# Base-21 Scrambling

Alexander Ivanov

***Abstract*—Base-2 scrambling is a well-known and well-proven technique widely adopted in modern communications. On the other hand, novel tasks such as the linguistic multiplexing case related to the 100BASE-X physical layer, necessitate scrambling using a non-binary base. In this paper, we seek to describe how to scramble on a base different from 2, the least prime number, solving a problem where base-21 scrambling is needed.**

## The Source of the Call

OUR PROBLEM originates from the linguistically multiplexed coding means we meet with in [1] and will refer to further in this paper as the design. That design practices the same 21 permissible, distinct, five-letter-long images[1] both before and after the scrambling, see Table III. Because the number 21 has no factor of 2 but a couple of 3 and 7 instead, the design necessitates a special scrambling means.

The design operates over the 100BASE-X physical medium dependent sublayer that leverages the FO-PMD per ISO/IEC 9314-3: 1990 to interface with glass optic fiber media, or else the TP-PMD per ANSI X3.263-1995 to interface with twisted pair media. The FO-PMD embodies no scrambling while the TP-PMD implements such a means. That means consists of a base-2 side-stream scrambler as the (pseudo) random number generator, followed by a base-2 cipher scrambler as the data stream bit-by-bit scrambling function.

In the paper, we assume the TP-PMD embedded scrambling means bypassed during operation of the design, but consider its random number generator either directly, as the randomity source for the proposed base-21 scrambling approach we call conservative, see Table I, or just referentially, as the original scheme for the progressive one,[2] see Table V.

A manuscript of this work was submitted to IEEE Communications Letters November 26, 2022 and rejected as not being in the scope of the journal.

Please sorry for the author has no time to find this work a new home, peer reviewed or not, except of arXiv, and just hopes there it meets its reader, one or maybe various, whom the author beforehand thanks for their regard.

A. Ivanov is with JSC Continuum, Yaroslavl, the Russian Federation.

[1]Those images are aliased (ordered) $x\mathrm{T}y$, where $x$ and $y$ are the patterns of "jumps" occurring in the earlier two letters (indexed $5n+0$) and the latter three letters (indexed $5n+2$), respectively, within a word. The permissible images are of $x \in \{1;2;3\}$ and $y \in \{1;2;3;4;5;6;7\}$. A "jump" (J) forces the line state is changing, instead of a "keep" (K) that relaxes the line state is remaining the same. The transport letters J and K comprise the transport alphabet all the $2^5 = 32$ possible images are constructed on. Restricting the coding scheme, via introducing the reduced transport dictionary consisting of the $3 \times 7 = 21$ permissible images only, prevents any run of more than three consecutive "keeps" in the plain stream, and the same in the cipher stream, when the introduced base-21 scrambling means is employed, see [1].

[2]In this paper, we focus on the conservative approach only.

TABLE I
Proposed Conservative Approach

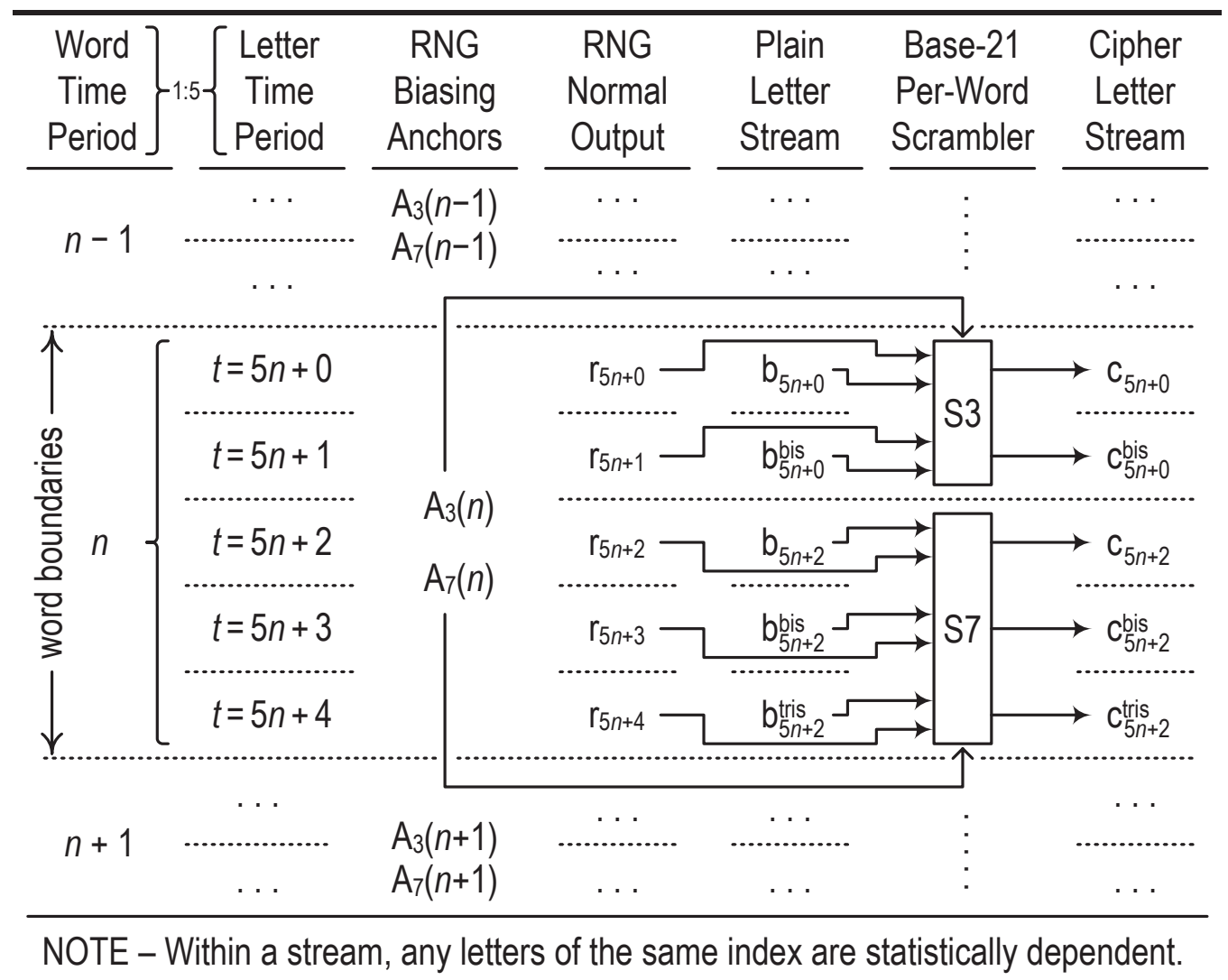

## Random Number Generator

TP-PMD [2] embodies a side-stream scrambler[3] generating five random bits per a 4B/5B coded block, i.e., one random bit per one coded bit sent. Thence, we say we employ a generator sourcing—not less than but not more than, thus, exactly—one random binary value per every letter.

We assume that a series of those values has no correlation (it is as weak as negligible) and therefore consider the values as statistically independent and refer to the corresponding bits sourced at the generator output as independent.

Within every word time period $(n)$ we form two groups of two and three random binary values the following manner:

$$\langle\ r_{5n+0} \quad r_{5n+1}\ \rangle \qquad \langle\ r_{5n+2} \quad r_{5n+3} \quad r_{5n+4}\ \rangle$$

where $5n+0$ to $5n+4$ are the letter time periods $(t)$ during which those values are generated.

The earlier group, set over the periods $5n+0$ and $5n+1$, gives $2^2 = 4$ distinct random values whose probabilities are distributed uniformly and equal to $p(2^2) = 1/4$. Similarly, the latter group, set over the periods $5n+2$, $5n+3$, and $5n+4$, gives $2^3 = 8$ distinct random values whose probabilities are distributed uniformly, too, but equal to $p(2^3) = 1/8$.

We use these grouped random values further in the random number generation process to set up the necessary ones.

[3]Typically implemented as a linear feedback shift register (LFSR); uses the generating polynomial $1 + x^9 + x^{11}$; ensures an average run of approximately two consecutive "keeps" and a maximum run of approximately 60 consecutive "keeps" in the cipher stream, see [2].

TABLE II
EXAMPLE BINARY-CODED BASE-PRIME SCRAMBLING

| Prime Number, $P$ → | $P = 2 = 2^1$ | $2^1 < P = 3 < 2^2$ | $2^2 < P = 5 < 2^3$ | $2^2 < P = 7 < 2^3$ |
|---|---|---|---|---|
| Numeral Code | direct binary | ternary, binary-coded into $2^2$ | quinary, binary-coded into $2^3$ | septenary, binary-coded into $2^3$ |
| Values Employed<br>[exc. values are deprecated] | 0; 1 — 2 out of 2 = $2^1$<br>(no excessive values) | 1; 2; 3 — 3 out of 4 = $2^2$<br>(value 0 [1 out of 4] is excessive) | 3; 4; 5; 6; 7 — 5 out of 8 = $2^3$<br>(values 0; 1; 2 [3 out of 8] are excessive) | 1; 2; 3; 4; 5; 6; 7 — 7 out of 8 = $2^3$<br>(value 0 [1 out of 8] is excessive) |
| Random Bit(s) Occupied | single | two, independent — $r_{x+0}$, $r_{x+1}$ | three, independent — $r_{x+0}$, $r_{x+1}$, $r_{x+2}$ | three, independent — $r_{x+0}$, $r_{x+1}$, $r_{x+2}$ |
| Scramling Bit(s) Provided | the same | two, dependent — $s_x$, $s_x^{bis}$ | three, dependent — $s_x$, $s_x^{bis}$, $s_x^{tris}$ | three, dependent — $s_x$, $s_x^{bis}$, $s_x^{tris}$ |
| NOTE – Bit indices are relative, assuming the ref bit position labeled "x".<br>Base-$P$ Random Number Generator (RNG) Function tabular-defined<br>↑ transparent for $P$ = 2, or anchor-based for $P$ > 2<br>↓ "advance counterclockwise" | Input = Out \| Prob.<br>0 = 0 \| $p(0)$=<br>1 = 1 \| $p(1)$=<br>Random Bit \| =1/2 (equal for 0 and 1)<br>NOTE – No anchor is needed.<br>NOTE – $P$ = 2 is the basis for the next primes. | In, $r_X$ \| Output, $s_X$ \| Prob.<br>0 0 \| 1 2 3 \| $p(1)$=<br>1 0 \| 2 3 1 \| $p(2)$=<br>0 1 \| 3 1 2 \| $p(3)$=<br>1 1 \| 1 2 3 \| =1/3<br>+0 +1 \| =1 =2 =3 \| =4/12 (equal for any # = 1,2,3)<br>Rnd Bits \| On Anchor State, $A_3(n)$<br>NOTE – Output is shown in decimal. | Input, $r_X$ \| Output, $s_X$ \| Prob.<br>0 0 0 \| 3 4 5 6 7 \| $p(3)$=<br>1 0 0 \| 4 5 6 7 3 \| $p(4)$=<br>0 1 0 \| 5 6 7 3 4 \| $p(5)$=<br>1 1 0 \| 6 7 3 4 5 \| $p(6)$=<br>0 0 1 \| 7 3 4 5 6 \| $p(7)$=<br>1 0 1 \| 3 4 5 6 7 \| =1/5<br>0 1 1 \| 4 5 6 7 3 \| =8/40<br>1 1 1 \| 5 6 7 3 4 \| (equal for any # = 3…7)<br>+0 +1 +2 \| =3 =4 =5 =6 =7<br>Bits \| On Anchor State, $A_5(n)$ | Input, $r_X$ \| Output, $s_X$ \| Probability<br>0 0 0 \| 1 2 3 4 5 6 7 \| $p(1)$=1/7<br>1 0 0 \| 2 3 4 5 6 7 1 \| $p(2)$=1/7<br>0 1 0 \| 3 4 5 6 7 1 2 \| $p(3)$=1/7<br>1 1 0 \| 4 5 6 7 1 2 3 \| $p(4)$=1/7<br>0 0 1 \| 5 6 7 1 2 3 4 \| $p(5)$=1/7<br>1 0 1 \| 6 7 1 2 3 4 5 \| $p(6)$=1/7<br>0 1 1 \| 7 1 2 3 4 5 6 \| $p(7)$=1/7<br>1 1 1 \| 1 2 3 4 5 6 7 \| (equal for any #)<br>+0 +1 +2 \| =1 =2 =3 =4 =5 =6 =7 \| for # = 1…7,<br>Bits \| On Anchor State, $A_7(n)$ \| $p(\#)$=8/56 |
| "Add" Operation | x ⊕ y = z | x ⊕ y = z | x ⊕ y = z | x ⊕ y = z |
| "Sub" Operation<br>↑ "advance clockwise" | z ⊖ y = x | z ⊖ y = x | z ⊖ y = x | z ⊖ y = x |
| Operations' Properties (mandatory) | time 1, time 2 = $P$<br>x ⊕ y ⊕ y = x<br>x ⊖ y ⊖ y = x<br>x ⊕ y ⊖ y = x<br>x ⊖ y ⊕ y = x<br>anyone, opposite | time 1, time 2, time 3 = $P$<br>x ⊕ y ⊕ y ⊕ y = x<br>x ⊖ y ⊖ y ⊖ y = x<br>x ⊕ y ⊖ y = x<br>x ⊖ y ⊕ y = x<br>anyone, opposite | time 1 … time 5 = $P$<br>x ⊕ y ⊕ y ⊕ y ⊕ y ⊕ y = x<br>x ⊖ y ⊖ y ⊖ y ⊖ y ⊖ y = x<br>x ⊕ y ⊖ y = x<br>x ⊖ y ⊕ y = x<br>anyone, opposite | time 1 … time 7 = $P$<br>x ⊕ y ⊕ y ⊕ y ⊕ y ⊕ y ⊕ y ⊕ y = x<br>x ⊖ y ⊖ y ⊖ y ⊖ y ⊖ y ⊖ y ⊖ y = x<br>x ⊕ y ⊖ y = x<br>x ⊖ y ⊕ y = x<br>anyone, opposite |
| Operations' Implementation | bit-wise XOR | binary-coded into $2^k$, modulo-$P$ over modulo-$2^k$, arithmetic addition/subtraction with conditional extra $2^k - P > 0$ on wraparounds | | |
| "Add", x + y → z | z ← x xor y | $z \leftarrow (0 + x + u - \alpha) \bmod 2^k + \Delta \cdot [(0 + x + u - \alpha) \operatorname{div} 2^k] - \beta$ | where | $u = (y + \gamma + \beta - \alpha) \bmod 2^k + \Delta \cdot [(y + \gamma + \beta - \alpha) \operatorname{div} 2^k]$ |
| "Sub", z − y → x | x ← z xor y | $x \leftarrow (2^k + x - u + \beta) \bmod 2^k + \Delta \cdot [(2^k + x - u + \beta) \operatorname{div} 2^k] - \beta$ | | $\alpha + \beta = \Delta = 2^k - P > 0$, $\Omega - \alpha = P - 1$, $\alpha \le \gamma \le \Omega$ |
| Constants $2^k$, $P$, $\Delta$ | $2^k = 2$, $P = 2$, $\Delta = 0$ | $2^k = 4$, $P = 3$, $\Delta = 1$ | $2^k = 8$, $P = 5$, $\Delta = 3$ | $2^k = 8$, $P = 7$, $\Delta = 1$ |
| Parameters α,β,γ,Ω | $\alpha = \beta = \gamma = 0$, $\Omega = 1$ | $\alpha = 1$, $\beta = 0$, $\gamma = 2$, $\Omega = 3$ | $\alpha = 3$, $\beta = 0$, $\gamma = 6$, $\Omega = 7$ | $\alpha = 1$, $\beta = 0$, $\gamma = 2$, $\Omega = 7$ |
| In × In × Out Space | {0;1} × {0;1} × {0;1} | {1;2;3} × {1;2;3} × {1;2;3} | {3;4;5;6;7} × {3;4;5;6;7} × {3;4;5;6;7} | {1;2;3;4;5;6;7} × {1;2;3;4;5;6;7} × {1;2;3;4;5;6;7} |
| Operations' Truth Tables assuming some $P$-ary logic is used (Boolean for $P = 2$) | "Add"(x, y): x\\y \| 0 1<br>0 \| 0 1<br>1 \| 1 0<br>"Sub"(z, y): x\\y \| 0 1<br>0 \| 0 1<br>1 \| 1 0 | "Add"(x, y): x\\y \| 1 2 3<br>1 \| 2 3 1<br>2 \| 3 1 2<br>3 \| 1 2 3<br>"Sub"(z, y): x\\y \| 1 2 3<br>1 \| 3 2 1<br>2 \| 1 3 2<br>3 \| 2 1 3 | "Add"(x, y): x\\y \| 3 4 5 6 7<br>3 \| 6 7 3 4 5<br>4 \| 7 3 4 5 6<br>5 \| 3 4 5 6 7<br>6 \| 4 5 6 7 3<br>7 \| 5 6 7 3 4<br>"Sub"(z, y): z\\y \| 3 4 5 6 7<br>3 \| 5 4 3 7 6<br>4 \| 6 5 4 3 7<br>5 \| 7 6 5 4 3<br>6 \| 3 7 6 5 4<br>7 \| 4 3 7 6 5 | "Add"(x, y): x\\y \| 1 2 3 4 5 6 7<br>1 \| 2 3 4 5 6 7 1<br>2 \| 3 4 5 6 7 1 2<br>3 \| 4 5 6 7 1 2 3<br>4 \| 5 6 7 1 2 3 4<br>5 \| 6 7 1 2 3 4 5<br>6 \| 7 1 2 3 4 5 6<br>7 \| 1 2 3 4 5 6 7<br>"Sub"(z, y): z\\y \| 1 2 3 4 5 6 7<br>1 \| 7 6 5 4 3 2 1<br>2 \| 1 7 6 5 4 3 2<br>3 \| 2 1 7 6 5 4 3<br>4 \| 3 2 1 7 6 5 4<br>5 \| 4 3 2 1 7 6 5<br>6 \| 5 4 3 2 1 7 6<br>7 \| 6 5 4 3 2 1 7 |
| Operations' Check Tables | "Add"("Sub"[x,y],y): x\\y \| 0 1<br>0 \| 0 0<br>1 \| 1 1<br>"Sub"("Add"[x,y],y): z\\y \| 0 1<br>0 \| 0 0<br>1 \| 1 1 | "Add"("Sub"[x, y], y): x\\y \| 1 2 3<br>1 \| 1 1 1<br>2 \| 2 2 2<br>3 \| 3 3 3<br>"Sub"("Add"[z, y], y): z\\y \| 1 2 3<br>1 \| 1 1 1<br>2 \| 2 2 2<br>3 \| 3 3 3 | "Add"("Sub"[x, y], y): x\\y \| 3 4 5 6 7<br>3 \| 3 3 3 3 3<br>… \| … … … … …<br>"Sub"("Add"[z, y], y): z\\y \| 3 4 5 6 7<br>3 \| 3 3 3 3 3<br>… \| … … … … … | "Add"("Sub"[x, y], y)<br>… … … … … … … …<br>"Sub"("Add"[z, y], y)<br>… … … … … … … … |
| Base-$P$ Scrambler (S‹$P$›) Function symbol-defined | $x_P$ → ⊕ → $x_S$ (↑ $s_X$) \| $x_S$ → ⊖ → $x_P$ (↑ $s_X$) | $x_P$ → ⊕ → $x_S$ (↑ $s_X$) \| $x_S$ → ⊖ → $x_P$ (↑ $s_X$) | $x_P$ → ⊕ → $x_S$ (↑ $s_X$) \| $x_S$ → ⊖ → $x_P$ (↑ $s_X$) | $x_P$ → ⊕ → $x_S$ (↑ $s_X$) \| $x_S$ → ⊖ → $x_P$ (↑ $s_X$) |
| Scrambling Direction → | Anyway | P. to Cipher \| C. to Plain | Plain to Cipher \| Cipher to Plain | Plain to Cipher \| Cipher to Plain |

TABLE III
TRANSPORT DICTIONARY

| Alias | Serial Image | $f_J$ | $f_K$ | Purpose | Alias | Serial Image | $f_J$ | $f_K$ | Purpose | Alias | Serial Image | $f_J$ | $f_K$ | Purpose | Alias | Serial Image | $f_J$ | $f_K$ | Purpose |
|---|---|---|---|---|---|---|---|---|---|---|---|---|---|---|---|---|---|---|---|
| 0T0 | K K K K K | 0/5 | 5/5 | *forbidden* | 1T0 | K J K K K | 1/5 | 4/5 | *forbidden* | 2T0 | J K K K K | 1/5 | 4/5 | *forbidden* | 3T0 | J J K K K | 2/5 | 3/5 | *forbidden* |
| 0T1 | K K K K J | 1/5 | 4/5 | *forbidden* | 1T1 | K J K K J | 2/5 | 3/5 | permitted | 2T1 | J K K K J | 2/5 | 3/5 | permitted | 3T1 | J J K K J | 3/5 | 2/5 | permitted |
| 0T2 | K K K J K | 1/5 | 4/5 | *forbidden* | 1T2 | K J K J K | 2/5 | 3/5 | permitted | 2T2 | J K K J K | 2/5 | 3/5 | permitted | 3T2 | J J K J K | 3/5 | 2/5 | permitted |
| 0T3 | K K K J J | 2/5 | 3/5 | *forbidden* | 1T3 | K J K J J | 3/5 | 2/5 | permitted | 2T3 | J K K J J | 3/5 | 2/5 | permitted | 3T3 | J J K J J | 4/5 | 1/5 | permitted |
| 0T4 | K K J K K | 1/5 | 4/5 | *forbidden* | 1T4 | K J J K K | 2/5 | 3/5 | permitted | 2T4 | J K J K K | 2/5 | 3/5 | permitted | 3T4 | J J J K K | 3/5 | 2/5 | permitted |
| 0T5 | K K J K J | 2/5 | 3/5 | *forbidden* | 1T5 | K J J K J | 3/5 | 2/5 | permitted | 2T5 | J K J K J | 3/5 | 2/5 | permitted | 3T5 | J J J K J | 4/5 | 1/5 | permitted |
| 0T6 | K K J J K | 2/5 | 3/5 | *forbidden* | 1T6 | K J J J K | 3/5 | 2/5 | permitted | 2T6 | J K J J K | 3/5 | 2/5 | permitted | 3T6 | J J J J K | 4/5 | 1/5 | permitted |
| 0T7 | K K J J J | 3/5 | 2/5 | *forbidden* | 1T7 | K J J J J | 4/5 | 1/5 | permitted | 2T7 | J K J J J | 4/5 | 1/5 | permitted | 3T7 | J J J J J | 5/5 | 0/5 | permitted |

NOTE – Among the permitted, the probability of occurrence of the letters J ("jump") and K ("keep") are $p_J \approx 0.61$ and $p_K \approx 0.39$, or about three and two per a word, respectively.

TABLE IV
EXAMPLE BINARY-CODED BASE-PRIME SCRAMBLERS

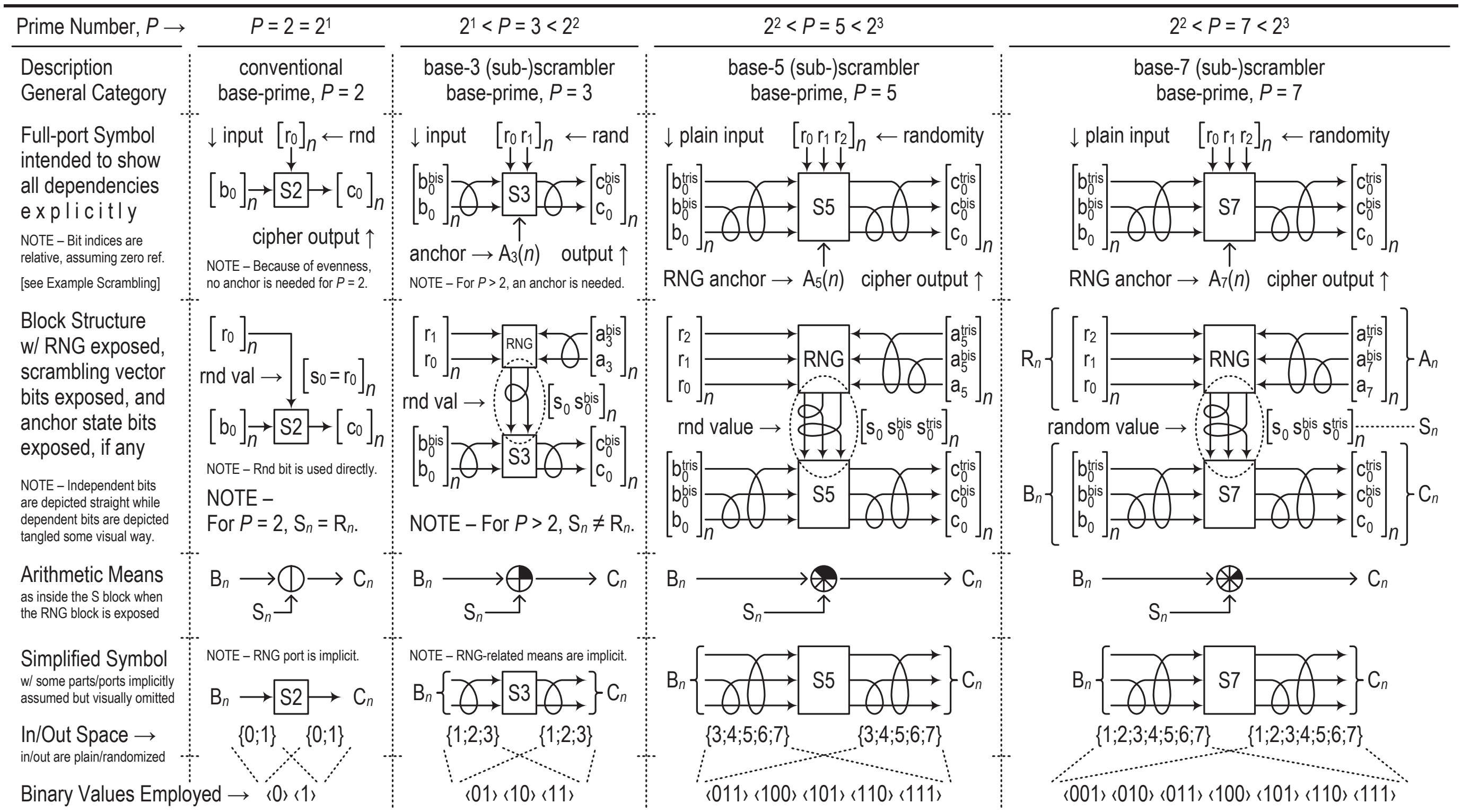

TABLE V
SUPPOSED PROGRESSIVE APPROACH

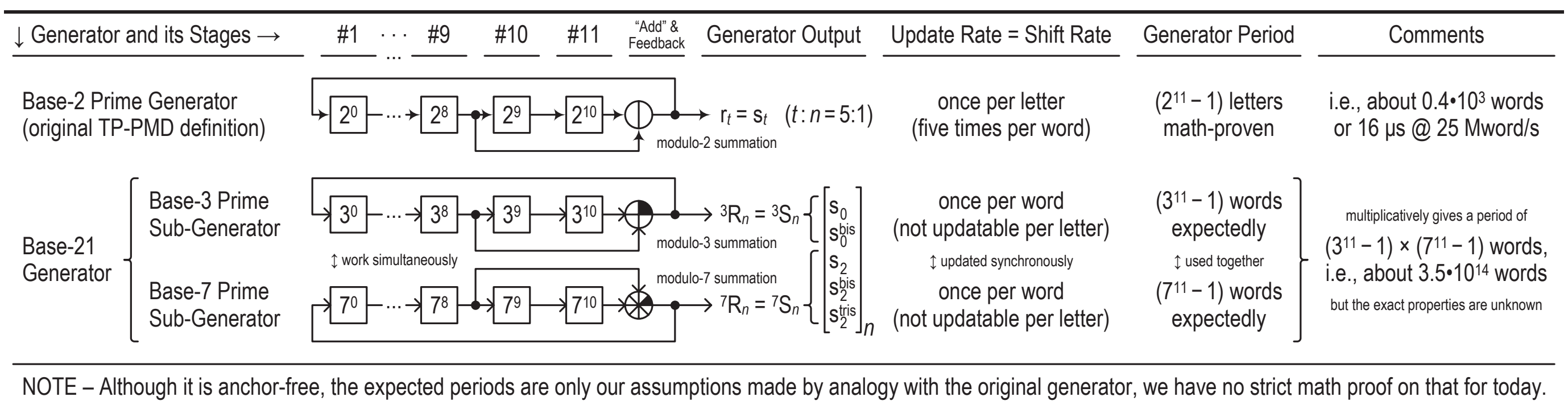

## RNG-BIASING ANCHORS

We introduce two cyclically running, continuously acting counters with the periods of 3 and 7, and refer to them as the anchors, $A_3(n)$ and $A_7(n)$, respectively, corresponding with the earlier and latter groups of the independent random bits sourced out of the generator, respectively.

Each anchor advances once a word time period so there is no state repetition found in a run of a length equal to the anchor period, i.e., 3 and 7 words, respectively. This gives us two independent, continuous series of periodically occurring values whose probabilities estimated over the corresponding period are distributed uniformly and equal to $p(3) = 1/3$ and $p(7) = 1/7$, respectively, see Table VI.

Each anchor immediately biases the corresponding grouped random value, i.e., $A_3(n)$ biases the grouped random value of $p(2^2) = 1/4$ to produce a random value of $p(3) = 1/3$ over every run of three word time periods, while $A_7(n)$ biases the grouped random value of $p(2^3) = 1/8$ to produce a random one of $p(7) = 1/7$ over every run of seven word time periods, the following manner:

$$|\Sigma|_3 \quad \frac{\begin{matrix} r_{5n+0} & r_{5n+1} \\ a_{5n+0} & a^{\text{bis}}_{5n+0} \end{matrix}}{\begin{matrix} s_{5n+0} & s^{\text{bis}}_{5n+0} \end{matrix}} \qquad \frac{\begin{matrix} r_{5n+2} & r_{5n+3} & r_{5n+4} \\ a_{5n+2} & a^{\text{bis}}_{5n+2} & a^{\text{tris}}_{5n+2} \end{matrix}}{\begin{matrix} s_{5n+2} & s^{\text{bis}}_{5n+2} & s^{\text{tris}}_{5n+2} \end{matrix}} \quad |\Sigma|_7$$

where $|\Sigma|_P$ is a modulo-$P$ summation, $a$'s indexed $5n+0$ and $5n+2$ all are the current values of the state bits of $A_3(n)$ and $A_7(n)$, respectively, and then $s$'s indexed $5n+0$ and $5n+2$ are the current values of the resulting random bits of the earlier and latter scrambling groups, respectively.

Hereupon, we can use such two resulting random values of

TABLE VI
RUNNING ANCHORS

| Word Time Period, $n$ mod 21 → | 0 | 1 | 2 | 3 | 4 | 5 | 6 | 7 | 8 | 9 | 10 | 11 | 12 | 13 | 14 | 15 | 16 | 17 | 18 | 19 | 20 |
|---|---|---|---|---|---|---|---|---|---|---|---|---|---|---|---|---|---|---|---|---|---|
| Base-3 Sub-Scrambler Anchor State, $A_3(n)$ | 1 | 2 | 3 | 1 | 2 | 3 | 1 | 2 | 3 | 1 | 2 | 3 | 1 | 2 | 3 | 1 | 2 | 3 | 1 | 2 | 3 |
| Base-7 Sub-Scrambler Anchor State, $A_7(n)$ | 1 | 2 | 3 | 4 | 5 | 6 | 7 | 1 | 2 | 3 | 4 | 5 | 6 | 7 | 1 | 2 | 3 | 4 | 5 | 6 | 7 |

NOTE – During normal transmission, this pattern is continuously repeated every LCM(3,7) = 21 words, where LCM(a,b) is the least common multiple of the numbers a and b.

TABLE VII
SCRAMBLER SYNCHRONIZATION SIGNALING

| Word Time Period → | ··· | intermediate | intermediate | $21M+0$ | $21M+1$ | ······· | $21M+19$ | $21M+20$ | intermediate | intermediate | ··· |
|---|---|---|---|---|---|---|---|---|---|---|---|
| Synchronization Phase w/in a Synchronization Cycle | ··· | inter-sync gap | sync preamble | sync word #1 | sync word #2 | ······· | sync word #20 | sync word #21 | sync postamble | inter-sync gap | ··· |
| Phase Duration | ··· | variable | fixed | fixed | fixed | ······· | fixed | fixed | fixed | variable | ··· |
| RNG and Anchors' Status | ··· | freeze | freeze | advance | advance | ······· | advance | advance | freeze | freeze | ··· |
| Anchors' States, $(A_3, A_7)$ | ··· | reset | (1,1) | (1,1) | (2,2) | ······· | (2,6) | (3,7) | (3,7) | reset | ··· |
| Stream Content Coded | ··· | forced K's | line ON pattern | $S_{21M+0}$ | $S_{21M+1}$ | ······· | $S_{21M+19}$ | $S_{21M+20}$ | line OFF pattern | forced K's | ··· |
| Stream Letters Sent | ··· | [K] | [J] K | any | any | ······· | any | any | [...] K | [K] | ··· |
| Line (Physical Media) State | ··· | zero | varies | varies | varies | ······· | varies | varies | varies to zero | zero | ··· |
| Visual Representation | ······· | | | | | ······· | | | | | ······· |
| | | | | Cycle $M$ | | | | | | $T_N = 0$ | |
| Link Startup Procedure Visual Representation | ··· ······ | $M$ ... | $T_{FEF}$ $M$ ... | ··· $M-1$ | $T_{F/S}$ | $M$ | $T_{F/S}$ $M+1$ | ··· | last of $M$ | $n = 0$ → | ··· |
| Inter-sync Gap Duration and Duration Meaning | any | $T_{FEF}$, signals a Far End Fault to the remote side | | $T_{FEF} >> T_F > T_S > 0$, signals the local de-scrambler is free or synchronized | | | | | | $T_N = 0$ | |
| Completeness Conditions | n/a | letter periods / word boundaries are detectable | | letter clocks are synchronized, word boundaries are detected and traced | | | | | | no sync cycles / gaps present | |
| Completeness Criterion | auto | link partner sends properly coded sync cycles | | link partner is active, local and remote de-scramblers are synchronized | | | | | | | |
| Transmission State → ··· | Reset | Link Partner Expectation | | Scrambler Synchronization | | | | | | Normal | ··· |

$p(3) = 1/3$ and $p(7) = 1/7$, respectively, in the scrambling process directly as the necessary random ones. Note that now the bit values in a group are statistically dependent, therefore we refer to the corresponding bits as dependent.

## CIPHER SCRAMBLER

Finally, we employ one base-3 prime sub-scrambler and one base-7 prime sub-scrambler, see Tables II[4] and IV, operating together, simultaneously and in parallel, the following manner, respectively, in the plain to cipher direction:

$$|+|_3 \quad \begin{array}{cc} b_{5n+0} & b^{\text{bis}}_{5n+0} \\ s_{5n+0} & s^{\text{bis}}_{5n+0} \\ \hline c_{5n+0} & c^{\text{bis}}_{5n+0} \end{array} \qquad \begin{array}{ccc} b_{5n+2} & b^{\text{bis}}_{5n+2} & b^{\text{tris}}_{5n+2} \\ s_{5n+2} & s^{\text{bis}}_{5n+2} & s^{\text{tris}}_{5n+2} \\ \hline c_{5n+2} & c^{\text{bis}}_{5n+2} & c^{\text{tris}}_{5n+2} \end{array} \quad |+|_7$$

and in the opposite, cipher to plain direction:

$$|-|_3 \quad \begin{array}{cc} c_{5n+0} & c^{\text{bis}}_{5n+0} \\ s_{5n+0} & s^{\text{bis}}_{5n+0} \\ \hline b_{5n+0} & b^{\text{bis}}_{5n+0} \end{array} \qquad \begin{array}{ccc} c_{5n+2} & c^{\text{bis}}_{5n+2} & c^{\text{tris}}_{5n+2} \\ s_{5n+2} & s^{\text{bis}}_{5n+2} & s^{\text{tris}}_{5n+2} \\ \hline b_{5n+2} & b^{\text{bis}}_{5n+2} & b^{\text{tris}}_{5n+2} \end{array} \quad |-|_7$$

where $|+|_P$ and $|-|_P$ are modulo-$P$ (equivalently, base-$P$) "addition" and "subtraction" operations, respectively, $b$'s and $c$'s are the binary codes of the current word's plain and cipher letters, i.e., before and after the scrambling, respectively.

Given a prime number $P > 2$, we apply two complimentary but distinct actions we above referred to as the "addition" and "subtraction" operations intended for scrambling, respectively, in the forward (plain to cipher = scrambling itself) and in the backward (cipher to plain = de-scrambling) directions.

## LINK SYNCHRONIZATION MEANS

We note that the base-21 scrambling means described above necessitates a link startup procedure[5] intended, among others, to synchronize the states of the random number generators as well as the states of the RNG-biasing anchors, across the link partners, see Table VII. Since such a procedure is not something new for but widely used in modern communications, we further see no theoretical obstacles for such scrambling to be applicable, implementable, and then usable.

## REFERENCES

[1] A. Ivanov, "Improving on, optimizing of, and explaining the data coding means and event coding means multiplexed over the 100BASE-X PMD sublayer," not published yet.

[2] Information Technology — Fibre Distributed Data Interface (FDDI) — Token Ring Twisted Pair Physical Layer Medium Dependent (TP-PMD), ANSI Std X3.263-1995, also but less known as INCITS 263-1995.

[4]In this table, $\gamma$ is the so called harmony parameter. Strictly speaking, we use $\gamma = \gamma_\alpha =$ "Add"$(\alpha, \alpha)$. Its value depends on what the implementer finds harmonic, e.g., $0+0=0$ or $0+0=1$, $1+1=1$ or $1+1=2$, etc. Also, varying the parameters $\alpha$, $\beta$, $\gamma$, and $\Omega$, the implementer can obtain the most suitable coding scheme, depending on the design goals.

[5]In spite of [1], where the scrambling site is placed at the Physical Coding Sublayer (PCS), here we consider the Physical Medium Attachment (PMA) sublayer the most suitable place for, assuming that exactly the PMA sublayer is responsible for many non-pure-digital functions, such as clock generation and recovery, letter synchronization and word alignment. So, we assume the link startup procedure is also a function of the PMA sublayer.

# Base-21 Word Alignment and Boundary Detection

Alexander Ivanov

***Abstract*—Word alignment is a common technique necessary in a serial data transmission system based on a means serializing a sequence of words into a stream of letters. In the heart of word alignment lies boundary detection, a basic technique intended to reliably separate words back, within a stream of letters. In this paper, we consider a useful—in the scope of these techniques—property of the base-21 words comprising the reduced transport dictionary employed in the linguistic multiplexing case related to the 100BASE-X physical layer.**

***Index Terms*—Ethernet, base-21 implicit comma, base-21 word alignment, base-21 word boundary detection, 100BASE-X.**

## Introduction

BASE-21 scrambling, as a particular case of a generalized base-prime scrambling, especially scrambling on a base different than a power of two, considered in [1], enables for a respective coding means to use the same five-letter-long serial images—that express the corresponding transport words[1] in a continuous text being serialized into a stream of letters—both before and after such scrambling is applied.

Thanks to this, the scrambled (or cipher) stream, as well as its (plain) source, preserves the statistical properties tied with the frequency of an expected shape—we will further refer to as an action—of the line state behavior, in a generalized form either of "jump" (J) or of "keep" (K), we observe at a selected position (letter time period, $t$) in the text, see Table I.

Such an observable action and its expected frequency give us an easy but objective ground to construct, at least in theory, a probabilistic measure capable to identify, for a respectively restricted variant of text and with a certain degree of veracity, somewhat similar in its purpose to what is called a comma in a serial continuous communication system.

In the rest of the paper, we describe a way to implement a means responsible for an appropriate base-21 word alignment and boundary detection task, based on the spoken above.

## Acquisition Loop

We consider a single-word-long acquisition procedure—we further refer to as the loop—that we apply to the text during an appropriate time interval—we, respectively, further refer to as the observation time—to acquire the information necessary to estimate the probabilities we are interested in.

A manuscript of this work was submitted to IEEE Communications Letters November 26, 2022 and rejected as not being in the scope of the journal.

Please sorry for the author has no time to find this work a new home, peer reviewed or not, except of arXiv, and just hopes there it meets its reader, one or maybe various, whom the author beforehand thanks for their regard.

A. Ivanov is with JSC Continuum, Yaroslavl, the Russian Federation.

[1]Those are aliased $x\mathrm{T}y$, where $x \in \{1; 2; 3\}$ and $y \in \{1; 2; 3; 4; 5; 6; 7\}$ simultaneously, all together comprising the reduced transport dictionary.

TABLE I
BASIC TERMS

| Curr. Letter Period, $t = 5n$ | | +0 | +1 | +2 | +3 | +4 | ($n$ = current word period) |
|---|---|---|---|---|---|---|---|
| K per-letter probability of occurrence, $p_{K,t}$ | = | $\frac{7}{21}$ ; | $\frac{7}{21}$ ; | $\frac{9}{21}$ ; | $\frac{9}{21}$ ; | $\frac{9}{21}$ | ← { as observed over 21 word periods |
| | | ↑ | ↑ | ↑ | ↑ | ↑ | |
| K (KEEP) per-letter generation frequency | = | $\frac{1}{3}$ ; | $\frac{1}{3}$ ; | $\frac{3}{7}$ ; | $\frac{3}{7}$ ; | $\frac{3}{7}$ | ← { GCD(1,3) = 1 / LCM(3,7) = 21 |
| | | ↑ | ↑ | ↑ | ↑ | ↑ | |
| $t : n = 5 : 1$ | | K — | J | K — | K — | J | |
| | | | | K — | J — | K | |
| | | | | K — | J — | J | |
| Base-3 Sub-Scrambler Possible Output, $C_3(n)$ | | J — | K | J — | K — | K | Base-7 Sub-Scrambler Possible Output, $C_7(n)$ |
| | | | | J — | K — | J | |
| choice 1-of-3 once per word (*uniformly distributed*) | | | | J — | J — | K | choice 1-of-7 once per word (*uniformly distributed*) |
| | | J — | J | J — | J — | J | |
| | | ↓ | ↓ | ↓ | ↓ | ↓ | |
| J (JUMP) per-letter generation frequency | = | $\frac{2}{3}$ ; | $\frac{2}{3}$ ; | $\frac{4}{7}$ ; | $\frac{4}{7}$ ; | $\frac{4}{7}$ | ← { GCD(2,4) = 2 / LCM(3,7) = 21 |
| | | ↓ | ↓ | ↓ | ↓ | ↓ | |
| J per-letter probability of occurrence, $p_{J,t}$ | = | $\frac{14}{21}$ ; | $\frac{14}{21}$ ; | $\frac{12}{21}$ ; | $\frac{12}{21}$ ; | $\frac{12}{21}$ | ← { as observed over 21 word periods |
| NOTE – Nominal letter and word time periods are 8 and 5 × 8 = 40 ns, respectively. | | | | | | | |

TABLE II
PROPOSED APPROACH

| ↓ Params, Variant/Alias → | JJ | JK | KK | KJ |
|---|---|---|---|---|
| Probabilistic Trace Seed | J J J J J | J J K K K | K K K K K | K K J J J |
| Probabilistic Trace Type | pure jump | mixed | pure keep | mixed |
| Action Counters Needed | 5(J) | 5(J)+5(K) | 5(K) | 5(J)+5(K) |
| Word Boundary Detector | looped pattern with single-n-strong plateau and peak | | | |

The loop covers over all the consecutive letter time periods together comprising the observation time,[2] in a modulo-five way: its first period is associated with every first letter period within the text's portion corresponded to the time, its second period is associated with every second letter period, and so on, indexed $t = 5n + i$, where $i \in \{0; 1; 2; 3; 4\}$, respectively, or simply $i$ when the common part is omitted, and referred to as the $i$-th letter period in the loop, assuming the observation time begins exactly at a word time period boundary.

In its turn, the observation time covers an integral number of transport word periods, or five times the number when we speak of either letter periods or letters themselves, because it anyway begins at a letter time period boundary.

Associating the loop with a seed, see Table II, we generate a trellis-like structure, that is vertically infinite (open) as well as horizontally cyclic (closed or looped), growing it step-by-step and up-to-down iteratively, see Tables III, IV, V, and VI.

[2]Nominal transport letter (word) time period is 8 ($5 \times 8 = 40$) ns. A ratio of a period to the observation time determines the scale of the observation.

TABLE III
JUMP PROBABILISTIC TRACE

| Step | Observation Time | Per-Letter/even or Inter-Letter/odd Probability of Occurrence | Scale (SN / SD) | GCD | Plateau | Peak | Linear Δ | Power Δ |
|---|---|---|---|---|---|---|---|---|
| initial | $21T_w$ = 0.84 μs | seed (JJJJJ) = $\frac{14}{21}$ ; $\frac{14}{21}$ ; $\frac{12}{21}$ ; $\frac{12}{21}$ ; $\frac{12}{21}$ ; "comma" | × 1 / 1 | | | | | |
| | $T_w$ = 40 ns (single word period) | $\frac{2•7}{3•7}$ $\frac{2•7}{3•7}$ $\frac{2•2•3}{3•7}$ $\frac{2•2•3}{3•7}$ $\frac{2•2•3}{3•7}$ | × 1 / 1 | | | | | |
| JJ0/even | $SD \times T_w$ = 0.84 μs | ··· 7 7 6 6 6 | × $2^1 / 3^1 7^1$ | — | no single/strong plateau (X=X) or peak (X) | | | |
| | double-letter two-path composite products = | 7•6 7•7 7•6 6•6 6•6 | × $2^2 / 3^2 7^2$ | — | recognizable pattern, no common factors | | | |
| JJ1/odd | $SD \times T_w$ = 17.64 μs | ··· < $7^1•6^1$ < <u>$7^2$</u> > $7^1•6^1$ > $6^2$ == $6^2$ < ··· | × $2^2 / 3^2 7^2$ | — | $6^2$ | $7^2$ | +36 % | +2.7 dB |
| | = 0.84 μs | 8÷8 <u>9÷10</u> 8÷8 6÷7 == 6÷7 | × 1 / 21 | | $7_{max}$ | $9_{min}$ | +29 % | +2.2 dB |
| | @ approximation | $6•21^1 < 2^2 6^2 = 144 < 7•21^1 < \cdots < 9•21^1 < 196 = 2^2 7^2 < 10•21^1$ | $3^2 7^2 = 21•21^1$ | $21^1$ | [see Positive Odd-step Boundary Detection Pattern] | | | |
| | triple-letter four-path composite products = | $7^3•6^1$ $7^3•6^1$ $7^1•6^3$ $6^4$ $7^1•6^3$ | × $2^4 / 3^4 7^4$ | $2^1 3^1$ | recognizable pattern, reducible numerators | | | |
| JJ2/even | $SD \times T_w$ ≈ 2.5 ms | ··· < $7^3$ == $7^3$ > $7^1•6^2$ > <u>$6^3$</u> < $7^1•6^2$ < ··· | × $2^5 / 3^3 7^4$ | — | $7^3$ | $6^3$ | −37 % | −4.0 dB |
| | = 5.88 μs | 24÷25 = 24÷25 18÷19 <u>15÷16</u> 18÷19 | × 1 / 21•7 | | $24_{min}$ | $16_{max}$ | −33 % | −3.5 dB |
| | @ approximation | $15•21^2 < 2^5 6^3 < 16•21^2 < \cdots < 24•21^2 < 2^5 7^3 < 25•21^2$ | $3^3 7^4 = 7•21•21^2$ | $21^2$ | [see Negative Even-step Boundary Detection Pattern] | | | |
| | quad-letter eight-path composite products = | $7^4•6^2$ $7^6$ $7^4•6^2$ $7^1•6^5$ $7^1•6^5$ | × $2^{10} / 3^6 7^8$ | $7^1$ | recognizable pattern, reducible numerators | | | |
| JJ3/odd | $SD \times T_w$ ≈ 24.0 s | ··· < $7^3•6^2$ < <u>$7^5$</u> > $7^3•6^2$ > $6^5$ == $6^5$ < ··· | × $2^{10} / 3^6 7^7$ | — | $6^5$ | $7^5$ | +116 % | +6.7 dB |
| | = 17.64 μs | 9÷10 <u>12÷13</u> 9÷10 5÷6 == 5÷6 | × 1 / $21^2$ | | $6_{max}$ | $12_{min}$ | +100 % | +6.0 dB |
| | @ approximation | $5•21^4 7^1 < 2^{10} 6^5 < 6•21^4 7^1 < \cdots < 12•21^4 7^1 < 2^{10} 7^5 < 13•21^4 7^1$ | $3^6 7^7 = 21•21•21^4 7^1$ | $21^4 7^1$ | [see Positive Odd-step Boundary Detection Pattern] | | | |
| ..... | ................. | ... | ........ | ... | ... | ... | ........ | ........ |

Linear Δ = $\frac{\text{Peak} - \text{Plateau}}{\text{Plateau}} \times 100$ [%]; Power Δ = $20 \times \log_{10} \frac{\text{Peak}}{\text{Plateau}}$ [dB]

TABLE IV
MIXED JUMP-THEN-KEEP PROBABILISTIC TRACE

| Step | Observation Time | Per-Letter/even or Inter-Letter/odd Probability of Occurrence | Scale (SN / SD) | GCD | Plateau | Peak | Linear Δ | Power Δ |
|---|---|---|---|---|---|---|---|---|
| initial | $21T_w$ = 0.84 μs | seed (JJKKK) = $\frac{14}{21}$ ; $\frac{14}{21}$ ; $\frac{9}{21}$ ; $\frac{9}{21}$ ; $\frac{9}{21}$ ; "comma" | × 1 / 1 | — | given as observed over 21 word periods [see Basic Terms and Proposed Approach] | | | |
| | single-letter one-path or elementary sources = | $\frac{2•7}{3•7}$ $\frac{2•7}{3•7}$ $\frac{3•3}{3•7}$ $\frac{3•3}{3•7}$ $\frac{3•3}{3•7}$ | × 1 / 1 | — | no factors common for all the numerators and for the denominator simultaneously | | | |
| JK0/even | $SD \times T_w$ = 0.84 μs | ··· 14 14 9 9 9 | × $1^1 / 3^1 7^1$ | — | no single/strong plateau (X=X) or peak (X) | | | |
| | double-letter two-path composite products = | 14•9 14•14 14•9 9•9 9•9 | × $1^1 / 3^2 7^2$ | — | recognizable pattern, no common factors | | | |
| JK1/odd | $SD \times T_w$ = 17.64 μs | ··· < $14^1•9^1$ < <u>$14^2$</u> > $14^1•9^1$ > $9^2$ == $9^2$ < ··· | × $1^1 / 3^2 7^2$ | — | $9^2$ | $14^2$ | +142 % | +7.7 dB |
| | = 0.84 μs | 6÷6 <u>9÷10</u> 6÷6 4÷5 == 4÷5 | × 1 / 21 | | $5_{max}$ | $9_{min}$ | +80 % | +5.1 dB |
| | @ approximation | $4•21^1 < 9^2 = 81 < 5•21^1 < 6•21^1 < 9•21^1 < 196 = 14^2 < 10•21^1$ | $3^2 7^2 = 21•21^1$ | $21^1$ | [see Positive Odd-step Boundary Detection Pattern] | | | |
| | triple-letter four-path composite products = | $14^3•9^1$ $14^3•9^1$ $14^1•9^3$ $9^4$ $14^1•9^3$ | × $1^1 / 3^4 7^4$ | $3^2$ | recognizable pattern, reducible numerators | | | |
| JK2/even | $SD \times T_w$ ≈ 0.9 ms | ··· < $14^3$ == $14^3$ > $14^1•9^2$ > <u>$9^3$</u> < $14^1•9^2$ < ··· | × $1^1 / 3^2 7^4$ | — | $14^3$ | $9^3$ | −73 % | −11.5 dB |
| | = 5.88 μs | 18÷19 = 18÷19 7÷8 <u>4÷5</u> 7÷8 | × 1 / 21•7 | | $18_{min}$ | $5_{max}$ | −72 % | −11.1 dB |
| | @ approximation | $4•7^1 21^1 < 9^3 < 5•7^1 21^2 < \cdots < 18•7^1 21^2 < 14^3 < 19•7^1 21^2$ | $3^2 7^4 = 7•21•7^1 21^1$ | $7^1 21^1$ | [see Negative Even-step Boundary Detection Pattern] | | | |
| | quad-letter eight-path composite products = | $14^4•9^2$ $14^6$ $14^4•9^2$ $14^1•9^5$ $14^1•9^5$ | × $1^1 / 3^4 7^8$ | $2^1 7^1$ | recognizable pattern, reducible numerators | | | |
| JK3/odd | $SD \times T_w$ ≈ 2.7 s | ··· < $14^3•9^2$ < <u>$14^5$</u> > $14^3•9^2$ > $9^5$ == $9^5$ < ··· | × $2^1 / 3^4 7^7$ | — | $9^5$ | $14^5$ | +811 % | +19.2 dB |
| | = 41.16 μs | 3÷4 <u>16÷17</u> 6÷7 1÷2 == 1÷2 | × 1 / $21•7^2$ | | $2_{max}$ | $16_{min}$ | +700 % | +18.1 dB |
| | @ approximation | $1•7^1 21^3 < 2^1 9^5 < 2•7^1 21^2 < \cdots < 16•7^1 21^3 < 2^1 14^5 < 17•7^1 21^3$ | $3^4 7^7 = 7•7•21•7^1 21^3$ | $7^1 21^3$ | [see Positive Odd-step Boundary Detection Pattern] | | | |
| ..... | ................. | ... | ........ | ... | ... | ... | ........ | ........ |

TABLE V
KEEP PROBABILISTIC TRACE

| Step | Observation Time | Per-Letter$^{/even}$ or Inter-Letter$^{/odd}$ Probability of Occurrence | Scale ( $SN$ / $SD$ ) | GCD | Plateau | Peak | Linear Δ | Power Δ |
|---|---|---|---|---|---|---|---|---|
| initial | $21T_w$ = 0.84 μs | seed (KKKKK) = $\frac{7}{21}$ ; $\frac{7}{21}$ ; $\frac{9}{21}$ ; $\frac{9}{21}$ ; $\frac{9}{21}$ ( , "comma") | × 1 / 1 | — | given as observed over 21 word periods [see Basic Terms and Proposed Approach] | | | |
| | single-letter one-path or elementary sources = | $\frac{7}{3•7}$ $\frac{7}{3•7}$ $\frac{3•3}{3•7}$ $\frac{3•3}{3•7}$ $\frac{3•3}{3•7}$ | × 1 / 1 | — | no factors common for all the numerators and for the denominator simultaneously | | | |
| KK0$^{/even}$ | $SD \times T_w$ = 0.84 μs | · · · 7 7 9 9 9 | × $1^1 / 3^1 7^1$ | — | no single/strong plateau (X=X) or peak ($\underline{X}$) | | | |
| | double-letter two-path composite products = | 9•7 7•7 9•7 9•9 9•9 | × $1^1 / 3^2 7^2$ | — | recognizable pattern, no common factors | | | |
| KK1$^{/odd}$ | $SD \times T_w$ = 17.64 μs | · · · > $9^1•7^1$ > $\underline{7^2}$ < $9^1•7^1$ < $9^2$ ═ $9^2$ > · · · | × $1^1 / 3^2 7^2$ | — | $9^2$ | $7^2$ | −40 % | −4.4 dB |
| | = 2.52 μs | 9÷9 $\underline{7÷7}$ 9÷9 11÷12 = 11÷12 | × 1 / 21•3 | | $11_{min}$ | $7_{max}$ | −36 % | −3.9 dB |
| | @ approximation | $7^2 = 49 = 7•7^1 < 9^1•7^1 = 63 = 9•7^1 < 11•7^1 < 81 = 9^2 < 12•7^1$ | $3^2 7^2 = 3•21•7^1$ | $7^1$ | [see Negative Odd-step Boundary Detection Pattern] | | | |
| | triple-letter four-path composite products = | $9^1•7^3$ $9^1•7^3$ $9^3•7^1$ $9^4$ $9^3•7^1$ | × $1^1 / 3^4 7^4$ | $3^2$ | recognizable pattern, reducible numerators | | | |
| KK2$^{/even}$ | $SD \times T_w$ ≈ 0.9 ms | · · · > $7^3$ ═ $7^3$ < $9^2•7^1$ < $\underline{9^3}$ > $9^2•7^1$ > · · · | × $1^1 / 3^2 7^4$ | — | $7^3$ | $9^3$ | +113 % | +6.5 dB |
| | = 17.64 μs | 7÷7 ═ 7÷7 11÷12 14÷15 11÷12 | × 1 / $21^2$ | | $7_{max}$ | $14_{min}$ | +100 % | +6.0 dB |
| | @ approximation | $7^3 = 343 = 7•7^2 < \cdots < 567 < \cdots < 14•7^2 < 729 = 9^3 < 15•7^2$ | $3^2 7^4 = 21•21•7^2$ | $7^2$ | [see Positive Even-step Boundary Detection Pattern] | | | |
| | quad-letter eight-path composite products = | $9^2•7^4$ $7^6$ $9^2•7^4$ $9^5•7^1$ $9^5•7^1$ | × $1^1 / 3^4 7^8$ | $7^1$ | recognizable pattern, reducible numerators | | | |
| KK3$^{/odd}$ | $SD \times T_w$ ≈ 2.7 s | · · · > $9^2•7^3$ > $\underline{7^5}$ < $9^2•7^3$ < $9^5$ ═ $9^5$ > · · · | × $1^1 / 3^4 7^7$ | — | $9^5$ | $7^5$ | −72 % | −10.9 dB |
| | = 370.44 μs | 3÷4 2÷3 3÷4 8÷9 8÷9 | × 1 / $21^3$ | | $8_{min}$ | $3_{max}$ | −63 % | −8.5 dB |
| | @ approximation | $2•21^1 7^3 < 7^5 < 3•21^1 7^3 < \cdots < 8•21^1 7^3 < 9^5 < 9•21^1 7^3$ | $3^4 7^7 = 21•21•21•21^1 7^3$ | $21^1 7^3$ | [see Negative Odd-step Boundary Detection Pattern] | | | |
| . . . . . | . . . . . . . . . . . . . . . . | ... | . . . . . . . . | . . . | . . . | . . . | . . . . . . . . | . . . . . . . . |

TABLE VI
MIXED KEEP-THEN-JUMP PROBABILISTIC TRACE

| Step | Observation Time | Per-Letter$^{/even}$ or Inter-Letter$^{/odd}$ Probability of Occurrence | Scale ( $SN$ / $SD$ ) | GCD | Plateau | Peak | Linear Δ | Power Δ |
|---|---|---|---|---|---|---|---|---|
| initial | $21T_w$ = 0.84 μs | seed (KKJJJ) = $\frac{7}{21}$ ; $\frac{7}{21}$ ; $\frac{12}{21}$ ; $\frac{12}{21}$ ; $\frac{12}{21}$ ( , "comma") | × 1 / 1 | — | given as observed over 21 word periods [see Basic Terms and Proposed Approach] | | | |
| | single-letter one-path or elementary sources = | $\frac{7}{3•7}$ $\frac{7}{3•7}$ $\frac{2•2•3}{3•7}$ $\frac{2•2•3}{3•7}$ $\frac{2•2•3}{3•7}$ | × 1 / 1 | — | no factors common for all the numerators and for the denominator simultaneously | | | |
| KJ0$^{/even}$ | $SD \times T_w$ = 0.84 μs | · · · 7 7 12 12 12 | × $1^1 / 3^1 7^1$ | — | no single/strong plateau (X=X) or peak ($\underline{X}$) | | | |
| | double-letter two-path composite products = | 12•7 7•7 12•7 12•12 12•12 | × $1^1 / 3^2 7^2$ | — | recognizable pattern, no common factors | | | |
| KJ1$^{/odd}$ | $SD \times T_w$ = 17.76 μs | · · · > $12^1•7^1$ > $\underline{7^2}$ < $12^1•7^1$ < $12^2$ ═ $12^2$ > · · · | × $1^1 / 3^2 7^2$ | — | $12^2$ | $7^2$ | −70 % | −9.4 dB |
| | = 0.84 μs | 4÷4 $\underline{2÷3}$ 4÷4 6÷7 ═ 6÷7 | × 1 / 21 | | $6_{min}$ | $3_{max}$ | −50 % | −6.0 dB |
| | @ approximation | $2•21^1 < 7^2 = 49 < 3•21^1 < 4•21^1 < 6•21^1 < 144 = 12^2 < 7•21^1$ | $3^2 7^2 = 21•21^1$ | $21^1$ | [see Negative Odd-step Boundary Detection Pattern] | | | |
| | triple-letter four-path composite products = | $12^1•7^3$ $12^1•7^3$ $12^3•7^1$ $12^4$ $12^3•7^1$ | × $1^1 / 3^4 7^4$ | $2^2 3^1$ | recognizable pattern, reducible numerators | | | |
| KJ2$^{/even}$ | $SD \times T_w$ ≈ 2.5 ms | · · · > $7^3$ ═ $7^3$ < $12^2•7^1$ < $\underline{12^3}$ > $12^2•7^1$ > · · · | × $2^2 / 3^3 7^4$ | — | $7^3$ | $12^3$ | +404 % | +14.0 dB |
| | = 5.88 μs | 3÷4 ═ 3÷4 9÷10 $\underline{15÷16}$ 9÷10 | × 1 / 21•7 | | $4_{max}$ | $15_{min}$ | +275 % | +11.5 dB |
| | @ approximation | $3•21^2 < 2^2 7^3 < 4•21^2 < \cdots < 15•21^2 < 2^2 12^3 < 16•21^2$ | $3^3 7^4 = 7•21•21^2$ | $21^2$ | [see Positive Even-step Boundary Detection Pattern] | | | |
| | quad-letter eight-path composite products = | $12^2•7^4$ $7^6$ $12^2•7^4$ $12^5•7^1$ $12^5•7^1$ | × $2^4 / 3^6 7^8$ | $7^1$ | recognizable pattern, reducible numerators | | | |
| KJ3$^{/odd}$ | $SD \times T_w$ ≈ 24.0 s | · · · > $12^2•7^3$ > $\underline{7^5}$ < $12^2•7^3$ < $12^5$ ═ $12^5$ > · · · | × $2^4 / 3^6 7^7$ | — | $12^5$ | $7^5$ | −93 % | −23.4 dB |
| | = 123.48 μs | 4÷5 $\underline{1÷2}$ 4÷5 20÷21 = 20÷21 | × 1 / $21^2$•7 | | $20_{min}$ | $2_{max}$ | −90 % | −20.0 dB |
| | @ approximation | $1•21^4 < 2^4 7^5 < 2•21^4 < \cdots < 20•21^4 < 2^4 12^5 < 21•21^4$ | $3^6 7^7 = 7•21•21•21^4$ | $21^4$ | [see Negative Odd-step Boundary Detection Pattern] | | | |
| . . . . . | . . . . . . . . . . . . . . . . | ... | . . . . . . . . | . . . | . . . | . . . | . . . . . . . . | . . . . . . . . |

TABLE VII
EVEN-STEP BOUNDARY DETECTION PATTERN

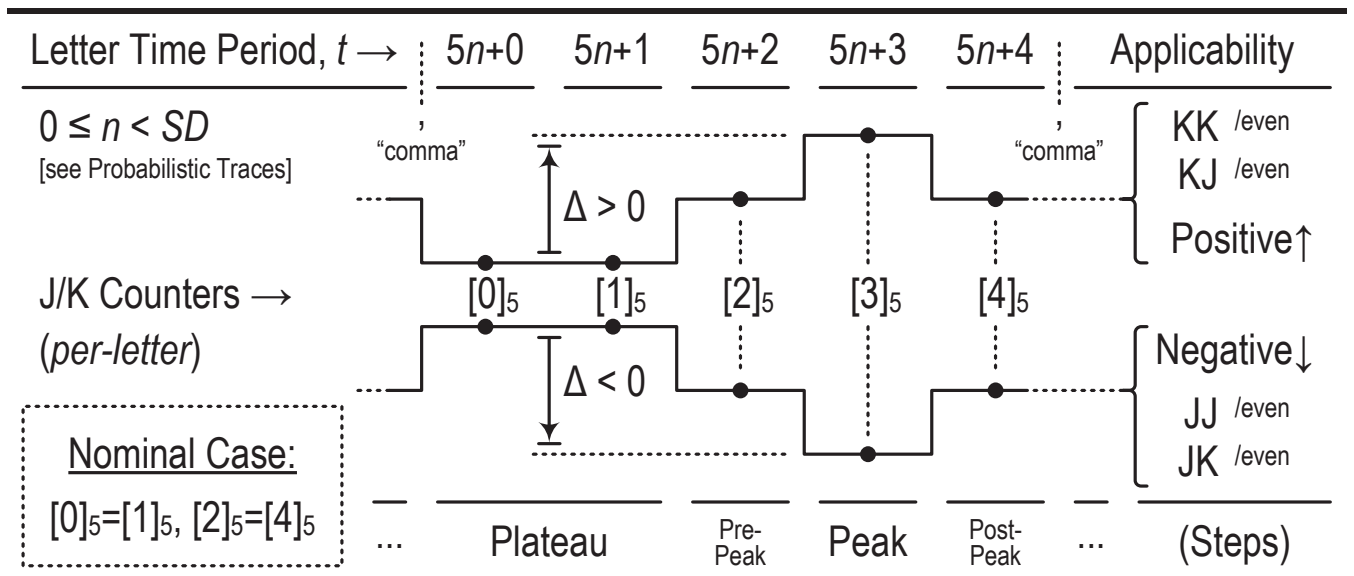

TABLE VIII
ODD-STEP BOUNDARY DETECTION PATTERN

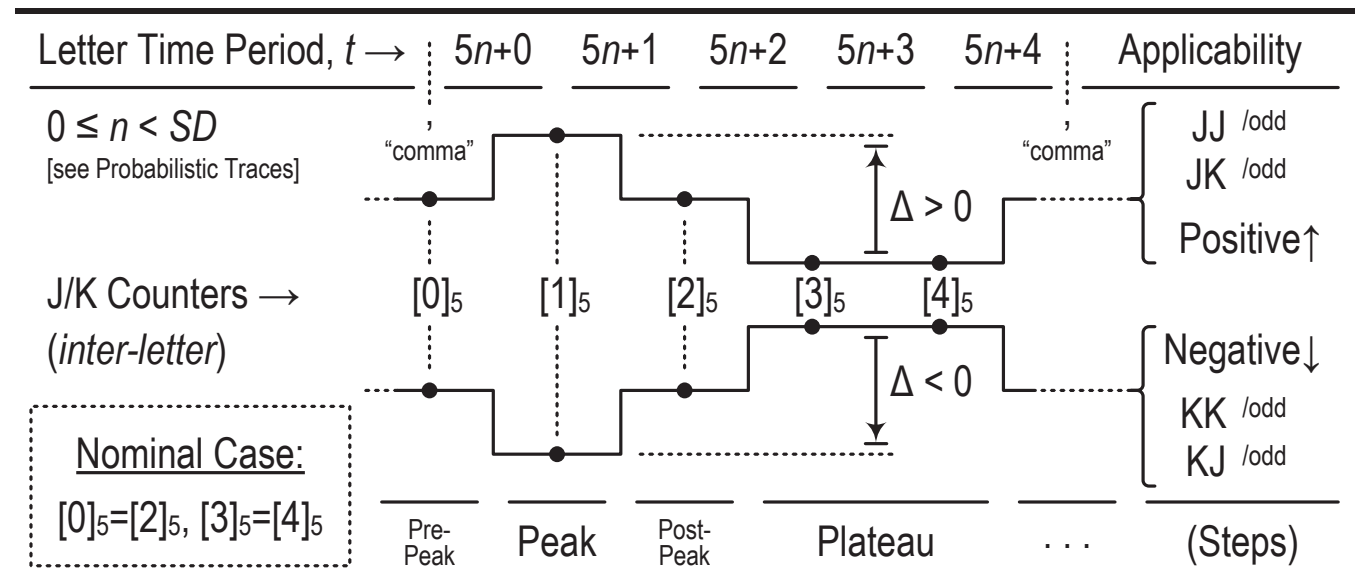

At each line corresponding to a step, odd or even, including the initial seed line, there are five nodes. Further, each node is designated with a number whose essentials are either just given, as for the nodes at the seed line, or, as for a node on a line below the seed, calculated so they are traceable up to the numbers of the nodes at the seed line, unambiguously.

A number of the $i$-th node at the seed line corresponds with the probability of an action, jump or keep, we expect to occur during the $i$-th letter period in the loop. Therefore, a number of a given node at a line below the seed is also corresponds with that probability, in a degree proportional to the number of distinguishable paths traveling up-to-down from the $i$-th node at the seed line into the given node at the given line.

An action perceivable by some node at the seed line, in the scope of that node, looks independent, or elementary, while at any line below the seed, it looks dependent because a node at such a line can perceive a composition of actions occurring, in the scope of that node, only and only all together, not any other way.[3] So, we read a node aliases for a respective action, elementary or composite, depending on where the node is, at the seed line or a line below the seed, respectively.

Thus, such a (node) number shows a (scaled) probability of occurrence of a respective action exactly at (even-step line) or right before (odd-step line) the $i$-th letter period in the loop, estimated during the observation time, forming up a point at a boundary detection pattern, see Tables VII and VIII.

Because of the spoken above, we refer to the whole trellis-like structure as a probabilistic trace of a seeded loop.

## ACTION COUNTERS

Since given a line in a probabilistic trace of a seeded loop, we associate each its node with a counter dedicated to count the number of times when a respective action occurs, during the corresponding letter periods, see Table II.

Until the observation time runs, a counter is reset. During the observation time, a counter can advance. After the observation time elapsed, a counter is stopped and held over. Each counter operates independently from other ones.

During the observation time runs, a counter associated with a node at the given line advances once per every respective action the node aliases for, in respect with the place of that node in the given trace, see Tables III, IV, V, and VI.

In this way, the value of a counter, as we acquired after the observation, holds the number of times we caught a respective action on the node the counter is associated with, i.e., such a value is proportional to the frequency of such an action taking its place relative to the $i$-th letter period in the loop.

Because of the spoken above, a value of an action counter provides us with a comparable measure we apply further.

## PATTERN MATCHING

Among the values provided by the counters associated with the nodes at the given line, we select a peak value and a plateau value, and then try to match the difference ($\Delta$) between these values as well as the relative interposition ($[i]_5$) of the counters showing these values, both at the same time, with the pattern corresponding to the line, see Tables VII and VIII, and finally compensate, by a cyclic rotation of the nodes at the line, for a modulo-five lag between the $i$-th letter period in the loop and an $i$-th letter period in the text, while the lag is not zero.

In the case of an even-step line, see Table VII, the boundary nearest to the peak is one full letter period after the $i$-th letter period in the loop, in which we observe the peak value.

In the case of an odd-step line, see Table VIII, the boundary nearest to the peak is a half letter period before what we could so call the accordant $i$-th letter period at the line.

This makes clear the boundaries of any word in the text we need to recover from the received stream of letters, and solves the problem we focus on in this paper, completely.

## CONCLUSION

Well, we considered a useful property intrinsic to a stream (or text) expressed in the base-21 transport words.

Being serialized, the base-21 words of the reduced transport dictionary statistically act (behave) themselves like a comma in a serial continuous communication system.[4]

Such “comma” seems implicit, or costless, because its use leads to no reduction in the payload of the stream, but gives a natural way for boundary detection.

It “separates” between every two consequitive words in the stream, allowing for word alignment, too.

## REFERENCES

[1] A. Ivanov, “Base-21 scrambling,” ································· ········ @*arXiv*, doi:10.48550/arXiv.yymm.nnnn (bundle), pp. 01–04.

[3] The scope of a node covers a set consisting of one (seed node) or more (other node) consecutive letter time periods within the observation time.

[4] Therefore we say that, instead of an explicit (symbolic) comma, the means considered in this paper is based on an implicit (statistic) comma.

# Quasi Base-21 Words

Alexander Ivanov

***Abstract*—In the paper, we introduce a new category of codes, inherited from, developed on, and then expanded beyond the so called base-21 words, and basically dive into them.**

***Index Terms*—Ethernet, linguistic multiplexing, base-21 words, base-21, quasi base-21 words, quasi base-21, QBTO.**

## Introduction

WRITTEN on a simple abstract transport alphabet consisting of just two letters, J and K, reflecting a line state change ("jump") and retain ("keep"), respectively, the base-21 words define the $3 \times 7 = 21$ distinct letter series aliased $x\mathrm{T}y$, where $x \in \{1, 2, 3\}$ and $y \in \{1, 2, 3, 4, 5, 6, 7\}$, comprising a code of interesting properties useful during data transmission, especially in scrambling [1] and alignment [2].[1]

Equating a code with (a set of) the words the code denotes, we mention the base-21 words as the reference code, whose length is $L = 5$ letters and capacity is $N = 21$ words, falling—not alone, but at least with its inverse replica, i.e., $y\mathrm{T}x$, of the same length and capacity, expectedly—in a category of codes each we could call exact base-21, or EBTO in short.

Based on the principles the reference code is designed with, we can construct a code of a longer length and, likely, a larger capacity—both in its absolute and relative values, i.e., per the whole length and per a letter of the length—in a shape aliased $x\mathrm{T}\cdots\mathrm{T}y$ or $y\mathrm{T}\cdots\mathrm{T}x$, that will fall in a category of new codes each we would call quasi base-21, or QBTO in short.

Comparing between codes, we indirectly refer to the reference code because compare between their bases expressing a sort of the relative value of the capacity, measured as:

$$\mathrm{BASE} = \mathrm{CAPACITY}^{\frac{5}{\mathrm{LENGTH}}},$$

supposing such the base equivalent, or quasi, where we speak about a quasi base-21 code, see Table I.

Like the reference one, any of the codes we consider in this paper consists of words such that—being issued in any order and quantity and then serialized—result in a stream preventing a run of more than three consecutive K's, a feature originating from the 4B/5B coding and, thus, usable in similar protocols, especially of Ethernet, like 100/1000BASE-X.[2]

Recalling the fate of submission of many prior works to the peer reviewed journal, such a try with this one also promises no chance, probably.

Please sorry for the author has no time to find this work a new home, peer reviewed or not, except of arXiv, and just hopes there it meets its reader, one or maybe various, whom the author beforehand thanks for their regard.

A. Ivanov is with JSC Continuum, Yaroslavl, the Russian Federation.

[1]In a broader sense, a base-<$z$> code, exact or quasi, may be built around any appropriate abstract alphabet declaring at least two letters.

[2]In a broader sense, a quasi base-21 code may be built around less or more stricter principles, compared to the reference, the base-21 words.

TABLE I
SHORT BRIEF

| Word Len. | Valid Images | Data Modulus | Most Binary Spaces | Eq. Base | Depiction |
|---|---|---|---|---|---|
| 5 (in letters) | 21 | $2^4$ | $2^4 + 2^2 + 2^0$ | $21_{\mathrm{exact}}$ | 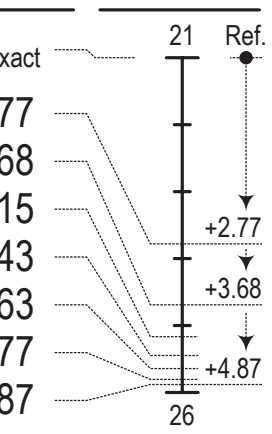  |
| 10 = 5 × 2 | 565 | $2^8$ | $2^9 + 2^5 + 2^4 + 2^2 + 2^0$ | 23.77 | |
| 15 = 5 × 3 | 15,033 | $2^{12}$ | $2^{13} + 2^{12} + 2^{11} + \ldots$ | 24.68 | |
| 20 = 5 × 4 | 400,025 | $2^{16}$ | $2^{18} + 2^{17} + 2^{12} + \ldots$ | 25.15 | |
| 25 = 5 × 5 | 10,644,589 | $2^{20}$ | $2^{23} + 2^{21} + 2^{17} + \ldots$ | 25.43 | |
| 30 ... | 283,250,477 | $2^{24}$ | $2^{28} + 2^{23} + 2^{22} + \ldots$ | 25.63 | |
| 35 ... | 7,537,241,009 | $2^{28}$ | $2^{32} + 2^{31} + 2^{30} + \ldots$ | 25.77 | |
| 40 ... | 200,564,541,425 | $2^{32}$ | $2^{37} + 2^{35} + 2^{34} + \ldots$ | 25.87 | |

## Capacity of a Code

Given a length of $L \geq 5$ letters, we construct a quasi base-21 code denoting a set of $N \geq 21$ words distinct in their serial images, when each of the images is expressed the same serial manner, whose number defines the (transport) capacity of such a code, assuming we exclude no image, that provides us with a code of the maximum capacity and, therefore, the maximum base possible for the given length, see Table I again.

However, when we exclude images, to reach some goal we have, e.g., to tune up a code to be comfortable in the use with a linguistic multiplexing process [3], we receive a quasi base-21 code of the same length, but of a capacity lower than before exclusion, by exact the number of excluded images, and thus of a proportionally decreased base, still perceiving it being of such a code despite of such a reduction, see Table II.[3]

## Balance of a Code

Given a code of the length $L$ and capacity $N$, we estimate its balance statistically, as a static in time, vector-like measure of the probability with which a selected letter, J or K, occurs at the $i$-th letter time period in the acquisition loop, during a potential transmission—all the words the given code denotes for participate in—assuming its duration infinite as well as its content random, contemporaneously, see Table III.[4]

When it is necessary, e.g., when it is in the goals we have, we actually change—that means that we additionally balance, completely rebalance, and even purposefully disbalance—the given code via exclusion of its images—everyone reflecting a distinct (sample of the) measure, with $i$-th value of either zero or $1/N$, exactly—by this receiving a new quasi base-21 code being else balanced while again still of such a code despite of a reduction from such a balancing, see Table IV.[5]

[3]Shown in this table for a given $L$, e.g., $L = 5$, is a code we could label shortly a $L(5)$ QBTO(20.00) code, hinting its major features for clear.

[4]Shown in this table is a $L(10)$ QBTO(23.77) code and the code we use as the ground for the material shown in the further tables of this paper.

[5]Shown in this table are a $L(10)$ QBTO(23.32), a $L(10)$ QBTO(23.00), and a $L(10)$ QBTO(22.63) codes, from the left to the right, respectively.

TABLE II
POSSIBLE CODING VARIANTS

| Length | Tr. Capacity | = $(2^U+1)\cdot 2^V$ | Eq. Base | Modulus | $N_C$ + $N_R$ = $N_E$ | GCD | $n_D$ = $n_e$ × $k$ | $E^{1/k}$ | Rest | Scr. Base(s) | Design Goal(s) |
|---|---|---|---|---|---|---|---|---|---|---|---|
| 5 × 1 | per word period 20 | = $(2^2+1)\cdot 2^2$ | $20_{exact}$ | $2^4=2^{4\times1}$ | 4 + 1 = 5 | $2^2$ | 8 = 4 × 2 | $2^1$ | $>2^6$ | dual — 2; 5 | mux simplicity |
| 5 × 2 | 544 | = $(2^4+1)\cdot 2^5$ | 23.32 | $2^8=2^{4\times2}$ | 8 + 9 = 17 | $2^5$ | — — — | — | $=2^0$ | dual — 2; 17 | performance |
| | 544 | = $(2^4+1)\cdot 2^5$ | 23.32 | $2^9=2^{4\times2+1}$ | 16 + 1 = 17 | $2^5$ | 48 = 12 × 4 | $2^1$ | $>2^{44}$ | dual — 2; 17 | protection |
| | 529 | not applicable | $23_{exact}$ | $2^8=2^{4\times2}$ | 16 + 7 = 23 | — | $8/2$ = $2/2$ × 4 | $2^1$ | $>2^4$ | single prime | balance |
| | $2^9=512$ | not applicable | 22.63 | $2^8=2^{4\times2}$ | 1 + 1 = 2 | $2^8$ | — — — | — | — | single binary | scr. simplicity |
| 5 × 3 | 13,824 | not applicable | $24_{exact}$ | $2^{12}=2^{4\times3}$ | 2 + 1 = 3 | $2^9$ | $2/3$ = $2/3$ × 1 | $2^1$ | $=2^0$ | dual — 2; 3 | performance |
| | 12,288 | = $(2^1+1)\cdot 2^{12}$ | 23.08 | $2^{13}=2^{4\times3+1}$ | 2 + 1 = 3 | $2^{12}$ | 2 = 2 × 1 | $2^1$ | $=2^0$ | dual — 2; 3 | prot. + mux |
| | 12,288 | = $(2^1+1)\cdot 2^{12}$ | 23.08 | $2^{12}=2^{4\times3}$ | 1 + 2 = 3 | $2^{12}$ | — — — | — | $=2^0$ | dual — 2; 3 | perf. + balance |
| | $2^{13}=8{,}192$ | not applicable | 20.16 | $2^{12}=2^{4\times3}$ | 1 + 1 = 2 | $2^{12}$ | — — — | — | — | single binary | scr. + balance |
| 5 × 4 | 393,216 | = $(2^1+1)\cdot 2^{17}$ | 25.04 | $2^{16}=2^{4\times4}$ | 1 + 2 = 3 | $2^{17}$ | — — — | — | $=2^0$ | dual — 2; 3 | performance |
| | 393,216 | = $(2^1+1)\cdot 2^{17}$ | 25.04 | $2^{18}=2^{4\times4+2}$ | 2 + 1 = 3 | $2^{17}$ | 2 = 2 × 1 | $2^1$ | $=2^0$ | dual — 2; 3 | prot. + mux |
| | 390,625 | not applicable | $25_{exact}$ | $2^{16}=2^{4\times4}$ | 4 + 1 = 5 | — | $8/8$ = $4/8$ × 2 | $2^1$ | $>2^6$ | single prime | balance + mux |
| | $2^{18}=262{,}144$ | not applicable | 22.63 | $2^{16}=2^{4\times4}$ | 1 + 3 = 4 | $2^{16}$ | — — — | — | $=2^1$ | single binary | perf.+scr.+bal. |
| 5 × 5 | 10,485,760 | = $(2^2+1)\cdot 2^{21}$ | 25.36 | $2^{23}=2^{4\times5+3}$ | 4 + 1 = 5 | $2^{21}$ | 8 = 4 × 2 | $2^1$ | $>2^6$ | dual — 2; 5 | prot. + mux |
| | $2^{23}=8{,}388{,}608$ | not applicable | 24.25 | $2^{20}=2^{4\times5}$ | 1 + 7 = 8 | $2^{20}$ | — — — | — | $>2^2$ | single binary | perf.+scr.+bal. |

TABLE III
BALANCING PRINCIPLES

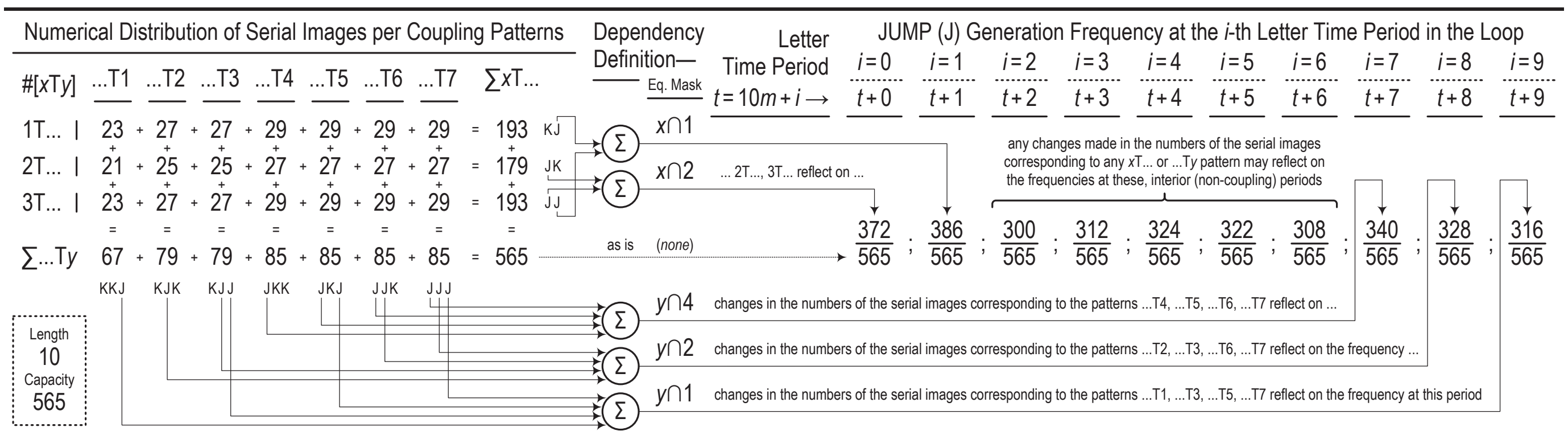

TABLE IV
BALANCING EXAMPLES

**565 ($23.77^2$) → 544 ($23.32^2$)**

| Row Descr. | #[xTy] | ...T1 | ...T2 | ...T3 | ...T4 | ...T5 | ...T6 | ...T7 | ∑xT... |
|---|---|---|---|---|---|---|---|---|---|
| Initial Distr. *including* NUMBERS TO CHANGE | 1T... | 23 | 27 | 27 | 29 | 29 | 29 | 29 | 193 |
| | 2T... | 21 | 25 | 25 | 27 | 27 | 27 | 27 | 179 |
| | 3T... | 23 | 27 | 27 | 29 | 29 | 29 | 29 | 193 |
| | ∑...Ty | 67 | 79 | 79 | 85 | 85 | 85 | 85 | 565 |
| Proposed Changes | 1T... | ▫ | ▫ | ▫ | −2 | ▫ | −12 | ▫ | −14 |
| | 2T... | ▫ | ▫ | ▫ | ▫ | ▫ | ▫ | ▫ | ▫ |
| | 3T... | ▫ | ▫ | ▫ | −7 | ▫ | ▫ | ▫ | −7 |
| | ∑...Ty | ▫ | ▫ | ▫ | −9 | ▫ | −12 | ▫ | −21 |
| Balancing Result | 1T... | 23 | 27 | 27 | 27 | 29 | 17 | 29 | 179 |
| | 2T... | 21 | 25 | 25 | 27 | 27 | 27 | 27 | 179 |
| | 3T... | 23 | 27 | 27 | 22 | 29 | 29 | 29 | 186 |
| | ∑...Ty | 67 | 79 | 79 | 76 | 85 | 73 | 85 | 544 |

| | t + 0 | t + 1 | t + 2 ••• t + 6 | t + 7 | t + 8 | t + 9 |
|---|---|---|---|---|---|---|
| JUMP (J) Gen. Freq. | 365/544 | 365/544 | 300 ~ 309 / 544 | 319/544 | 316/544 | 316/544 |
| $p_{JUMP}(t+i)$ | .67 | .67 | .55 ~ .57 | .59 | .58 | .58 |
| $p_{KEEP}(t+i)$ | .33 | .33 | .45 ~ .43 | .41 | .42 | .42 |

**565 ($23.77^2$) → 529 ($23^2$)**

| Row Descr. | #[xTy] | ...T1 | ...T2 | ...T3 | ...T4 | ...T5 | ...T6 | ...T7 | ∑xT... |
|---|---|---|---|---|---|---|---|---|---|
| Initial Distr. | 1T... | 23 | 27 | 27 | 29 | 29 | 29 | 29 | 193 |
| | 2T... | 21 | 25 | 25 | 27 | 27 | 27 | 27 | 179 |
| | 3T... | 23 | 27 | 27 | 29 | 29 | 29 | 29 | 193 |
| | ∑...Ty | 67 | 79 | 79 | 85 | 85 | 85 | 85 | 565 |
| Proposed Changes | 1T... | ▫ | −2 | −2 | −2 | −2 | −2 | −4 | −14 |
| | 2T... | ▫ | ▫ | ▫ | ▫ | ▫ | ▫ | ▫ | ▫ |
| | 3T... | ▫ | ▫ | ▫ | −12 | ▫ | −10 | ▫ | −22 |
| | ∑...Ty | ▫ | −2 | −2 | −14 | −2 | −12 | −4 | −36 |
| Balancing Result | 1T... | 23 | 25 | 25 | 27 | 27 | 27 | 25 | 179 |
| | 2T... | 21 | 25 | 25 | 27 | 27 | 27 | 27 | 179 |
| | 3T... | 23 | 27 | 27 | 17 | 29 | 19 | 29 | 171 |
| | ∑...Ty | 67 | 77 | 77 | 71 | 83 | 73 | 81 | 529 |

| | t + 0 | t + 1 | t + 2 ••• t + 6 | t + 7 | t + 8 | t + 9 |
|---|---|---|---|---|---|---|
| JUMP (J) Gen. Freq. | 350/529 | 350/529 | 298/529 | 308/529 | 308/529 | 308/529 |
| $p_{JUMP}(t+i)$ | .66 | .66 | .56 | .58 | .58 | .58 |
| $p_{KEEP}(t+i)$ | .34 | .34 | .44 | .42 | .42 | .42 |

**565 ($23.77^2$) → 512 ($22.63^2$)**

| Row Descr. | #[xTy] | ...T1 | ...T2 | ...T3 | ...T4 | ...T5 | ...T6 | ...T7 | ∑xT... |
|---|---|---|---|---|---|---|---|---|---|
| Initial Distr. | 1T... | 23 | 27 | 27 | 29 | 29 | 29 | 29 | 193 |
| | 2T... | 21 | 25 | 25 | 27 | 27 | 27 | 27 | 179 |
| | 3T... | 23 | 27 | 27 | 29 | 29 | 29 | 29 | 193 |
| | ∑...Ty | 67 | 79 | 79 | 85 | 85 | 85 | 85 | 565 |
| Proposed Changes | 1T... | −2 | −2 | −2 | −3 | −2 | −2 | −9 | −22 |
| | 2T... | ▫ | ▫ | ▫ | −1 | ▫ | ▫ | −7 | −8 |
| | 3T... | ▫ | ▫ | ▫ | −10 | ▫ | −12 | −1 | −23 |
| | ∑...Ty | −2 | −2 | −2 | −14 | −2 | −24 | −17 | −53 |
| Balancing Result | 1T... | 21 | 25 | 25 | 26 | 27 | 27 | 20 | 171 |
| | 2T... | 21 | 25 | 25 | 26 | 27 | 27 | 20 | 171 |
| | 3T... | 23 | 27 | 27 | 19 | 29 | 17 | 28 | 170 |
| | ∑...Ty | 65 | 77 | 77 | 71 | 83 | 71 | 68 | 512 |

| $t = 10m$ → | t + 0 | t + 1 | t + 2 ••• t + 6 | t + 7 | t + 8 | t + 9 |
|---|---|---|---|---|---|---|
| JUMP (J) Gen. Freq. | 341/512 | 341/512 | 290/512 | 293/512 | 293/512 | 293/512 |
| $p_{JUMP}(t+i)$ | .67 | .67 | .57 | .57 | .57 | .57 |
| $p_{KEEP}(t+i)$ | .33 | .33 | .43 | .43 | .43 | .43 |

TABLE V
IMAGE GENERATION RULES

| Input Sequence Lasts With | Body Letters Left (BLL) | Appendable Options (• = J or K) | Such Per Run | Output Sequence Lasts With | ENDEC Framework Element | Weight Pattern (body applicable only) $K^0$ $K^1$ $K^2$ $K^3$ |
|---|---|---|---|---|---|---|
| ⋯$\overline{\text{KKK}}$ [ new image... ] | all | • J | 2/3 | ⋯J | $K^{\le 2}$ 2•... $K^0$ or 2•... $K^0$ | |
| | | J K | 1/3 | ⋯J K | $K^{\le 2}$ 1•... $K^1$ or 1•... $K^1$ | |
| ⋯J | @ ≥3 | • • J | 4/8 | ⋯J | $K^0$ 4•... $K^0$ | 4 2 1 1 |
| | | • J K | 2/8 | ⋯J K | $K^0$ 2•... $K^1$ | |
| | | J K K | 1/8 | ⋯J K K | $K^0$ 1•... $K^2$ | |
| | | K K K | 1/8 | ⋯J K K K | $K^0$ 1•... $K^3$ | |
| | @ 2 | • J | 2/4 | ⋯J | $K^0$ 2•... $K^0$ | 2 1 1 — |
| | | J K | 1/4 | ⋯J K | $K^0$ 1•... $K^1$ | |
| | | K K | 1/4 | ⋯J K K | $K^0$ 1•... $K^2$ | |
| | @ 1 | J | 1/2 | ⋯J | $K^0$ 1•... $K^0$ | 1 1 — — |
| | | K | 1/2 | ⋯J K | $K^0$ 1•... $K^1$ | |
| ⋯J K | @ ≥2 | • J | 2/4 | ⋯J | $K^1$ 2•... $K^0$ | 2 1 — 1 |
| | | J K | 1/4 | ⋯J K | $K^1$ 1•... $K^1$ | |
| | | K K | 1/4 | ⋯J K K K | $K^1$ 1•... $K^3$ | |
| | @ 1 | J | 1/2 | ⋯J | $K^1$ 1•... $K^0$ | 1 — 1 — |
| | | K | 1/2 | ⋯J K K | $K^1$ 1•... $K^2$ | |
| ⋯J K K | @ ≥1 | J | 1/2 | ⋯J | $K^2$ 1•... $K^0$ | 1 — — 1 |
| | | K | 1/2 | ⋯J K K K | $K^2$ 1•... $K^3$ | |
| ⋯J K K K | @ ≥1 | J | 1/1 | ⋯J | $K^3$ → $K^0$ | 1 — — — |
| ⋯J | — | $\overline{\text{KKK}}$ | 7/7 | ⋯$\overline{\text{KKK}}$ [ ...done! ] | $K^0$ 7 $K^{\le 2}$ or $K^0$ 7 | |
| ⋯J K | — | $\overline{\text{KKK}}$ | 7/7 | | $K^1$ 7 $K^{\le 2}$ or $K^1$ 7 | |
| ⋯J K K | — | $\overline{\text{KK}}$ • | 6/6 | | $K^2$ 6 $K^{\le 2}$ or $K^2$ 6 | |
| ⋯J K K K | — | J • • | 4/4 | | $K^3$ 4 $K^{\le 2}$ or $K^3$ 4 | |

## FRAMEWORK OF A CODE

Given a base-21 code, quasi or exact and balanced or not, we assign each its distinct serial image a distinct word index, $0 \leq B < N$, reading distinct is among images and indices of all the words the given code denotes for, respectively.[6]

After the given code becomes so each its word relates with a distinct image as well as with a distinct index, we complete it with a framework, i.e., a formalized description intended to match between word indices and serial images, resolving for a given index with its corresponding image and vise versa, that is enough to unambiguously express the underlying laws used for encoding and decoding, contemporaneously.

Such a framework does implement the principles the code it describes is designed with, see Table V briefly.

Such a framework consists of a number of interlinked bins, where a bin itself links (a span of) indices, (a span of) letter periods, and (a choice of) image syllables, all together setting up the required matching, see Tables V and VI.

Such a framework allows for an appropriate coding means to encode and decode every issued word, $W(m)$, pointed by an index, i.e., expecting $W(m) = B$, into and from the image that word corresponds to, see Table VII.

[6]Other indices mentioned in this paper are the word (time period) index, $m \geq 0$, the letter (time period) index, $t \geq 0$, and the letter (time period) index in the loop, $0 \leq i < L$, interrelated as $t = L \cdot m + i$.

TABLE VI
BASE-23.77 ENDEC FRAMEWORK

| Head Periods | | Body (Letter Time) Periods | | | | | | Tail Periods | B |
|---|---|---|---|---|---|---|---|---|---|
| 2 • 193 | $K^0$ | 4 • 27 | $K^0$ | 2 • 7 | $K^0$ | | | 7 | 0 |
| | | | | | | | | 7 | 7 |
| | | | | 1 • 7 | $K^1$ | | | 7 | 14 |
| | | | | 1 • 6 | $K^2$ | | | 6 | 21 |
| | | | | 2 • 7 | $K^0$ | | | 7 | 27 |
| | | | | | | | | 7 | 34 |
| | | | | 1 • 7 | $K^1$ | | | 7 | 41 |
| | | | | 1 • 6 | $K^2$ | | | 6 | 48 |
| | | | | 2 • 7 | $K^0$ | | | 7 | 54 |
| | | | | | | | | 7 | 61 |
| | | | | 1 • 7 | $K^1$ | | | 7 | 68 |
| | | | | 1 • 6 | $K^2$ | | | 6 | 75 |
| | | | | 2 • 7 | $K^0$ | | | 7 | 81 |
| | | | | | | | | 7 | 88 |
| | | | | 1 • 7 | $K^1$ | | | 7 | 95 |
| | | | | 1 • 6 | $K^2$ | | | 6 | 102 |
| | | 2 • 25 | $K^1$ | 2 • 7 | $K^0$ | | | 7 | 108 |
| | | | | | | | | 7 | 115 |
| | | | | 1 • 6 | $K^1$ | | | 7 | 122 |
| | | | | 1 • 4 | $K^3$ | | | 4 | 129 |
| | | | | 2 • 7 | $K^0$ | | | 7 | 133 |
| | | | | | | | | 7 | 140 |
| | | | | 1 • 6 | $K^1$ | | | 7 | 147 |
| | | | | 1 • 4 | $K^3$ | | | 4 | 154 |
| | | 1 • 21 | $K^2$ | 1•14 | $K^0$ | 1•7 | $K^0$ | 7 | 158 |
| | | | | | | 1•7 | $K^1$ | 7 | 165 |
| | | | | 1•7 | $K^3$ | → | $K^0$ | 7 | 172 |
| | | 1 • 14 | $K^3$ | → | $K^0$ | 1•7 | $K^0$ | 7 | 179 |
| | | | | | | 1•7 | $K^1$ | 7 | 186 |
| | | 4 • 27 | $K^0$ | 2 • 7 | $K^0$ | | | 7 | 193 |
| | | | | | | | | 7 | 200 |
| | | | | | | | | | 207 |
| • • • | | | | | | | | | |
| | | 1 • 14 | $K^3$ | → | $K^0$ | 1•7 | $K^0$ | 7 | 372 |
| | | | | | | 1•7 | $K^1$ | 7 | 379 |
| 1 • 179 | $K^1$ | 2 • 52 | $K^0$ | 4 • 7 | $K^0$ | | | 7 | 386 |
| | | | | | | | | 7 | 393 |
| | | | | | | | | 7 | 400 |
| | | | | | | | | 7 | 407 |
| | | | | 2 • 7 | $K^1$ | | | 7 | 414 |
| | | | | | | | | 7 | 421 |
| | | | | 1 • 6 | $K^2$ | | | 6 | 428 |
| | | | | 1 • 4 | $K^3$ | | | 4 | 434 |
| | | | | 4 • 7 | $K^0$ | | | 7 | 438 |
| | | | | | | | | 7 | 445 |
| | | | | | | | | 7 | 452 |
| | | | | | | | | 7 | 459 |
| | | | | 2 • 7 | $K^1$ | | | 7 | 466 |
| | | | | | | | | 7 | 473 |
| | | | | 1 • 6 | $K^2$ | | | 6 | 480 |
| | | | | 1 • 4 | $K^3$ | | | 4 | 486 |
| | | 1 • 48 | $K^1$ | 2 • 14 | $K^0$ | 2•7 | $K^0$ | 7 | 490 |
| | | | | | | | | 7 | 497 |
| | | | | | | 2•7 | $K^1$ | 7 | 504 |
| | | | | | | | | 7 | 511 |
| | | | | 1 • 13 | $K^1$ | 1•7 | $K^0$ | 7 | 518 |
| | | | | | | 1•6 | $K^2$ | 6 | 525 |
| | | | | 1 • 7 | $K^3$ | → | $K^0$ | 7 | 531 |
| | | 1 • 27 | $K^3$ | → | $K^0$ | 2 • 7 | $K^0$ | 7 | 538 |
| | | | | | | | | 7 | 545 |
| | | | | | | 1 • 7 | $K^1$ | 7 | 552 |
| | | | | | | 1 • 6 | $K^2$ | 6 | 559 |
| | | | | | | | | | 565 |
| $t+0$ | $t+1$ | $t+2$ | $t+3$ | $t+4$ | $t+5$ | $t+6$ | $t+7$ | $t+8$ / $t+9$ | B |

Base-23.77 Encoder Decoder Framework Formula: 565 =
2 • { 4•(2•7+1•7+1•6) + 2•(2•7+1•7+1•4) + 1•[1•(1•7+1•7)+1•(1•7)] + 1•[1•(1•7+1•7)] } +
1 • { 2•(4•7+2•7+1•6+1•4) + 1•[2•(1•7+1•7)+1•(1•7+1•6)+1•(1•7)] + 1•[1•(2•7+1•7+1•6)] }

TABLE VII
EXAMPLE ENDEC PROCESS

| Step | Encoding (Serial Image Generation) Procedure |
|---|---|
| start | Given $W(m) = 455 \rightarrow rest_0 = 455$ |
| 1E | Head Run of 2 letters $\rightarrow 0 \le rest_0 < (386) \le rest_0 < (565) \rightarrow$ |
| | $weight_1 \cdot bin_1 = 1 \cdot 179$ // from 0≡386, incl., to 179≡565, excl. |
| | $rest_1 = (rest_0 - 386) \bmod bin_1 = 69$ |
| | $code_1 = (rest_0 - 386)$ div $bin_1 = 0$   $K^{\le 2}$ 1 • ... $K^1$ |
| | (a) $option_1 = HEAD(bin_1, code_1) = JK$ // t+0, t+1 |
| | $BLL_1 = 5$ // the next run is a body run |
| 2E | Body Run of $\Delta_2 = \Delta BLL(option_1, BLL_1) = 2$ letters $\rightarrow$ |
| | $0 \le rest_1 < (104) \le rest_1 < (152) \le rest_1 < (179) \rightarrow$ |
| | $weight_2 \cdot bin_2 = 2 \cdot 52$ // from 0≡386, incl., to 104≡490, excl. |
| | $rest_2 = (rest_1 - 0) \bmod bin_2 = 17$   $K^1$ 2 • ... $K^0$ |
| | $code_2 = (rest_1 - 0)$ div $bin_2 = 1$ |
| | (b) $option_2 = BODY(\Delta_2, code_2) = KJ$ // t+2, t+3 |
| | $BLL_2 = BLL_1 - \Delta_2 = 3$ // the next run is a body run |
| 3E | Body Run of $\Delta_3 = \Delta BLL(option_2, BLL_2) = 3$ letters $\rightarrow$ |
| | $0 \le rest_2 < (28) \le rest_2 < (42) \le rest_2 < (48) \le rest_2 < (52) \rightarrow$ |
| | $weight_3 \cdot bin_3 = 4 \cdot 7$ // from 0≡438, incl., to 28≡466, excl. |
| | $rest_3 = (rest_2 - 0) \bmod bin_3 = 3$   $K^0$ 4 • ... $K^0$ |
| | $code_3 = (rest_2 - 0)$ div $bin_2 = 2$ |
| | (c) $option_3 = BODY(\Delta_3, code_3) = KJJ$ // t+4, t+5, t+6 |
| | $BLL_3 = BLL_2 - \Delta_3 = 0$ // no body letters left |
| 4E | Tail Run of 3 letters $\rightarrow 0 \le rest_3 < (bin_3) \rightarrow$   $K^0$ 7 $K^{\le 2}$ |
| | $code_4 = rest_3 = 3$ |
| | (d) $option_4 = TAIL(code_4) = JKK$ // t+7, t+8, t+9 |
| done | Transmitted Image is JK KJ KJJ JKK |

| bin | code | HEAD |
|---|---|---|
| 193 | 0 | J J |
| 193 | 1 | K J |
| 179 | 0 (a) | J K |
| **Δ** | **code** | **BODY** |
| 3 | 0 | J J J |
| | 1 | J K J |
| | 2 (c) | K J J |
| | 3 | K K J |
| | 4 | J J K |
| | 5 | K J K |
| | 6 | J K K |
| | 7 | K K K |
| 2 | 0 | J J |
| | 1 (b) | K J |
| | 2 | J K |
| | 3 | K K |
| 1 | 0 | J |
| | 1 | K |
| | **code** | **TAIL** |
| | 0 | J J J |
| | 1 | J J K |
| | 2 | J K J |
| | 3 (d) | J K K |
| | 4 | K J J |
| | 5 | K J K |
| | 6 | K K J |

| Decoding (Word Index Restoration) Procedure | Step |
|---|---|
| Received Image is JK-KJKJJ-JKK | start |
| Head Run of 2 letters → JK-KJKJJ-JKK → $option_1 = JK$ → | 1D |
| signal an error if $HEAD^{-1}(option_1)$ does not exist, otherwise: | |
| (a) $\{ bin_1, code_1 \} = HEAD^{-1}(option_1) = \{ 179, 0 \}$ | |
| (a) $weight_1 \cdot bin_1 = 1 \cdot 179 \rightarrow (386) \le germ_1 < (565)$ | |
| (a) $germ_1 = 386 + bin_1 \times code_1 = 386 + 179 \times 0 = 386$ | |
| (a) $BLL_1 = 5$ | |
| Body Run of $\Delta_2 = \Delta BLL(option_1, BLL_1) = 2$ letters → | 2D |
| · · · · · · · · · · · · · · · · · JK-KJKJJ-JKK → $option_2 = KJ$ → | |
| (b) $\{ {}^{dummy}\Delta_2, code_2 \} = BODY^{-1}(option_2) = \{ 2, 1 \}$ | |
| (b) $weight_2 \cdot bin_2 = 2 \cdot 52 \rightarrow (386) \le germ_2 < (490)$ | |
| (b) $germ_2 = germ_1 + bin_2 \times code_2 = 386 + 52 \times 1 = 438$ | |
| (b) $BLL_2 = BLL_1 - \Delta_2 = 3$ | |
| Body Run of $\Delta_3 = \Delta BLL(option_2, BLL_2) = 3$ letters → | 3D |
| · · · · · · · · · · · · · · · · · JK-KJKJJ-JKK → $option_3 = KJJ$ → | |
| (c) $\{ {}^{dummy}\Delta_3, code_3 \} = BODY^{-1}(option_3) = \{ 3, 2 \}$ | |
| (c) $weight_3 \cdot bin_3 = 4 \cdot 7 \rightarrow (386{\equiv}438) \le germ_3 < (414{\equiv}466)$ | |
| (c) $germ_3 = germ_2 + bin_3 \times code_3 = 438 + 7 \times 2 = 452$ | |
| (c) $BLL_3 = BLL_2 - \Delta_3 = 0$ | |
| Tail Run of 3 letters → JK-KJKJJ-JKK → $option_4 = JKK$ → | 4D |
| signal an error if $TAIL^{-1}(option_4)$ does not exist, otherwise: | |
| (d) $code_4 = TAIL^{-1}(option_4) = 3$ | |
| (d) $(germ_3) \le germ_4 < (germ_3 + bin_3) \rightarrow (452) \le germ_4 < (459)$ | |
| (d) $germ_4 = germ_3 + 1 \times code_4 = 452 + 1 \times 3 = 455$ | |
| Found $W(m) = germ_4 \rightarrow W(m) = 455$ | done |

TABLE VIII
BALANCING INDEX CONVERSION

| Input / Argument | | Scrambled Space | | Output / Result |
|---|---|---|---|---|
| $0 \le W_{user}(m) < N_{user}$ → single linear segment without skips | scrambling | → $0 \le W_{scr}(m) < N_{scr}$ → single linear segment without skips | remapping | → $0 \le W_{endec}(m) < N$ many linear segments with $N - N_{scr}$ skips |
| NOTE – In a generalized case, $0 \le N_{user} \le N_{scr} \le N$ = available transport capacity. | | | | |

TABLE IX
SUMMARY OUTLINE

| ↓Prop., Eq. Base→ | 23.77 | 24.68 | 25.15 | 25.40 | 25.43 | • • • | 28.72 |
|---|---|---|---|---|---|---|---|
| Letters per Word | 10 | 15 | 20 | ∞ | 25 | • • • | ∞ |
| Bits per "Nibble" | 4.57 | 4.63 | 4.65 | 4⅔ (average) | 4.67 | • • • | 4.84 (approx.) |
| Bins in Formula | 41 | 378 | 3,540 | LETTER STUFFING | 33,120 | • • • | theoretical LIMIT |
| Body Runs | 2÷3 | 4÷6 | 5÷9 | | 7÷12 | • • • | |

## APPLICATION OF A CODE

Given a strawman of a coding means, we consider a bundle of related coding spaces, that usually consists of a user coding space, a scrambling space, and an ENDEC space, each actual at the respective stage of the coding, see Table VIII.

A user coding space, or a user space, features a capacity of a shape like $(2^{\mathrm{U}}+1)\cdot 2^{\mathrm{V}}$, typically, see Table II again, where the larger part $2^{\mathrm{U}}\cdot 2^{\mathrm{V}}$ responds for a data quantity transmission while the smaller part $2^{\mathrm{V}}$ responds for a control transmission, respectively, as well as the parts together are represented as a single, continuous span of word indices, $0 \le B < N_{\text{user}}$.

A scrambling space features a capacity equal to a power of two, in the best case, a power of a prime higher than two, in a typical case, or a multiple of some (powers of some) primes, in a worst case we always try to avoid, see Table II yet again, habitually represented as a single, continuous span of indices, too, but anyway not less than of the user, $N_{\text{scr}} \ge N_{\text{user}}$.

An ENDEC space features a capacity equal to the capacity of either the underlying line code, or the code driving it, that is a quasi base-21 code in the scope of this paper, but matched with the capacity of the scrambling by index skips responding to word exclusions, how necessary, ensuring $N \ge N_{\text{scr}}$.

## CONCLUSION

Designing a coding means, we usually conduct a search for the best code, that results in a selection of appropriate options we further will make our single choice just across.[7]

Ranging between, we can estimate the performance of each option, bounded due to the principles staying behind it, as so:

$$\frac{\text{BITS}}{\text{“NIBBLE”}} = \log_2 \text{BASE} < 5\,,$$

reachable at the cost of framework bins, that's fair to all quasi base-21 words we mentioned earlier, see Table IX.[8]

## REFERENCES

[1] A. Ivanov, "Base-21 scrambling," · · · · · · · · · · · · · · · · · · · · · · · · · · · · · · · · · · · · · · · · · *@arXiv*, doi:10.48550/arXiv.yymm.nnnn (bundle), pp. 01–04.

[2] A. Ivanov, "Base-21 word alignment and boundary detection," · · · · · · · · · · · · · · · · *@arXiv*, doi:10.48550/arXiv.yymm.nnnn (bundle), pp. 05–08.

[3] A. Ivanov, (this is the initial topic in the whole work about) "Data coding means and event coding means multiplexed over the 100BASE-X PMD sublayer," not published yet.

[7] Generalizing, we can treat an exact base-21 code as a quasi base-21 code and such is true while allowed, however the opposite is not, anyway.

[8] A single term label for such a code could also be a $(N \uparrow 5/L = z)$ code, e.g., $(565 \uparrow 5/10 = 23.77)$, or similar, e.g., $(565^{5/10} = 23.77)$.

# Quasi Base-21 Words Generated Compactly

Alexander Ivanov

***Abstract*—In this paper, we continue to consider the newborn category of codes—so called quasi base-21 words, set QBTO in short—inherited from and then expanded beyond the progenitor as well as their root, so called exact base-21 words, set EBTO in short. Codes of the new category are still abstract, completely as their progenitor, but, as their progenitor, too, demonstrate useful deterministic features and shapeful probabilistic properties, both helpful in running a plain line code that performs not so perfect alone in the respective application.**

***Index Terms*—Ethernet, ENDEC framework, framework, quasi base-21 words, quasi base-21 code, quasi base-21, QBTO.**

## INTRODUCTION

ORIGINATING from the so called base-21 words, known for their interesting properties related to scrambling [1] and alignment [2], quasi base-21 words [3] describe a broader category of codes, which inherit the principles the progenitor was designed with, as well as its transport alphabet.

Referentially equated with the (set of) words it denotes for, a quasi base-21 code is a) characterized by its length, $L$, and capacity, $N$, measured in letters and words, respectively, then b) ranged by its base, $z = N^{5/L}$, and performance, $\log_2 z$, and finally c) described by its ENDEC framework [3].

A framework of a code consists of a number of so looking like bins, that increases dramatically along an increase in the length of such a code, rendering an implementation of such a code impractical as well as the avalanching complexity of the underlying coding means unacceptable, expectedly.

In the rest of this paper, we consider a way enabling for us to describe such a framework very compactly, that, in its turn, makes the respective code, as an integral part of the respective quasi base-21 coding means in the case, applicable in modern communication protocols, especially like of Ethernet.

## PLOT OF A CODE

Given a code,[1] we describe a plot of that code, constructing such a plot the following way, based on the framework.

Similarly to [3], we connect a bin of such a framework with a certain time interval, in the whole word time, pointed by the letter index, $i$, during which syllables of that bin occur.

TABLE I
COMPACT $L = 5$ ENDEC FRAMEWORK EXAMPLE

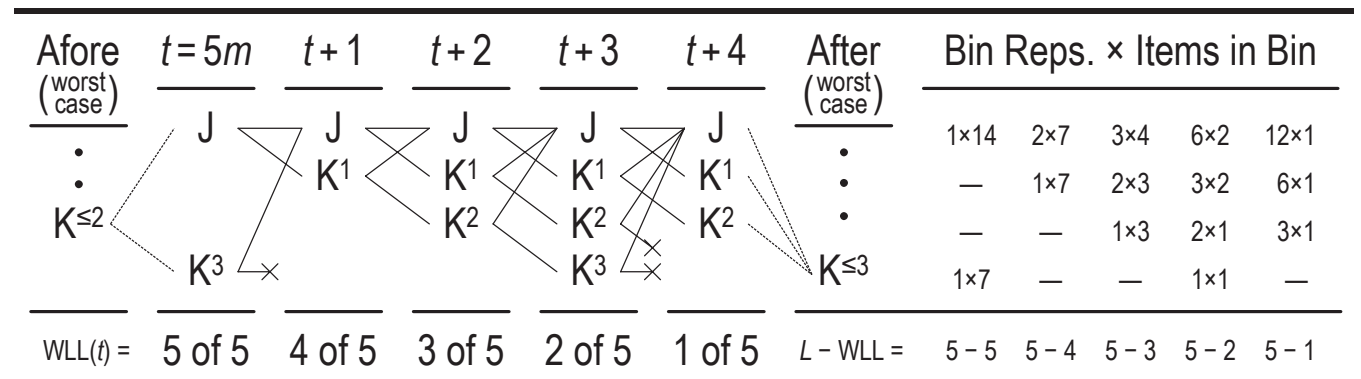

TABLE II
EXAMPLE DETAILS

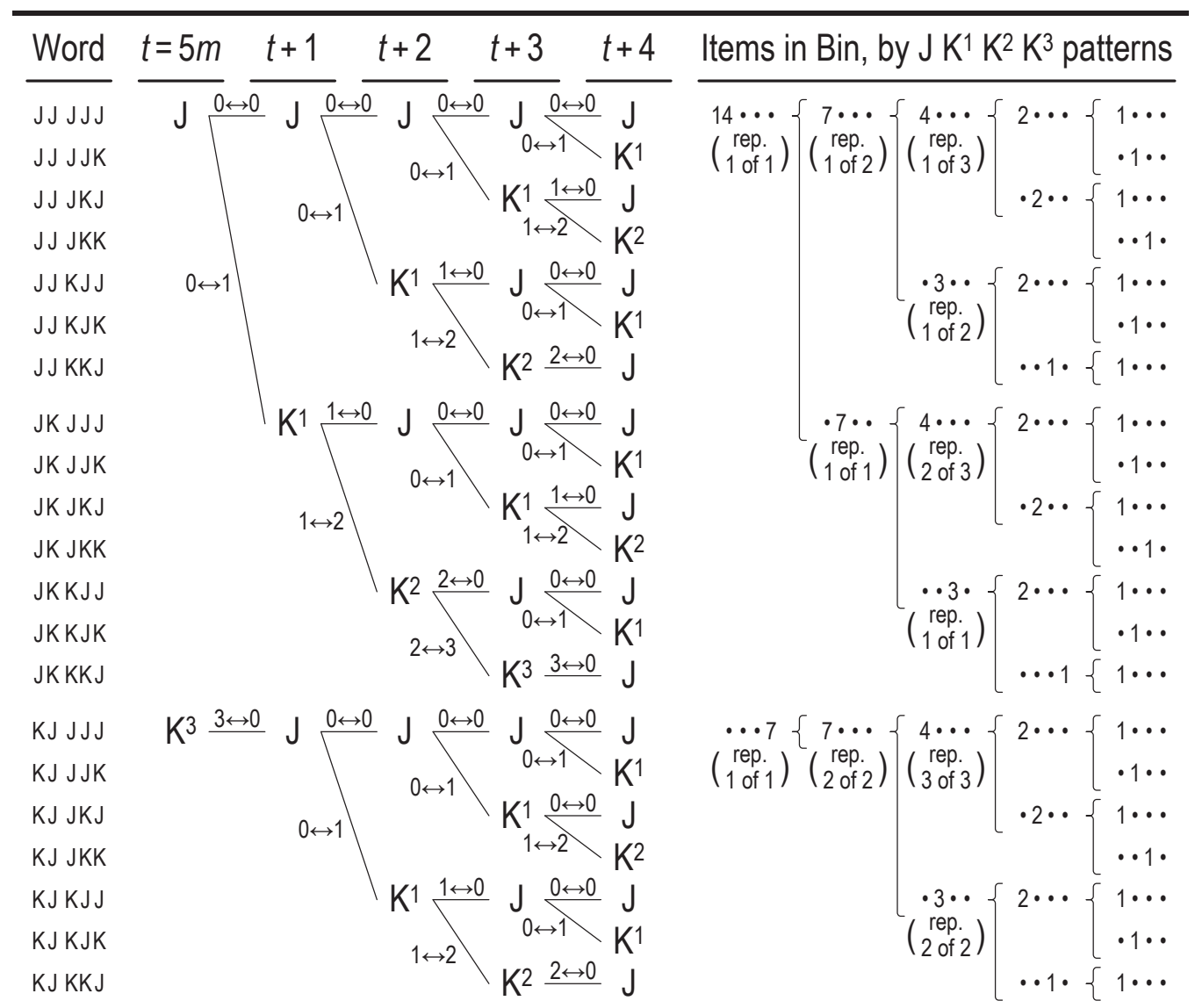

TABLE III
ALLOWED J–K TRANSITS

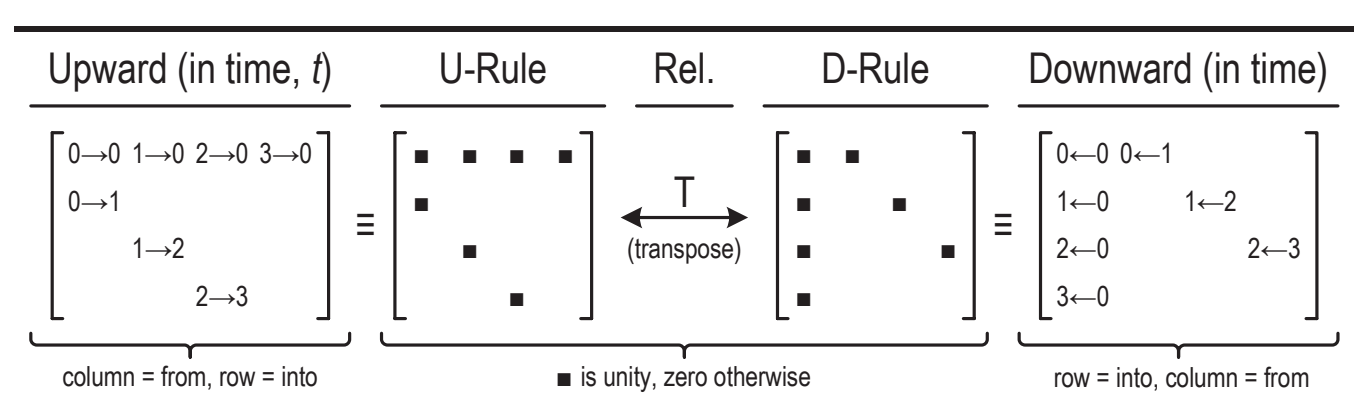

Oppositely to [3], we shorten the number of syllables a bin describes as well as the number of letters a syllable envelops to just one and one, respectively, further dealing right with so unified bins and their repetitions only, see Table I.[2]

Recalling the fate of submission of many prior works to the peer reviewed journal, such a try with this one also promises no chance, probably.

Please sorry for the author has no time to find this work a new home, peer reviewed or not, except of arXiv, and just hopes there it meets its reader, one or maybe various, whom the author beforehand thanks for their regard.

A. Ivanov is with JSC Continuum, Yaroslavl, the Russian Federation.

[1]In this paper, we consider only unbalanced exact and quasi base-21 codes, i.e., exact and quasi base-21 codes with no word exclusions, see [3].

[2]In this paper, we consider an exact base-21 code as a quasi-base-21 code, assuming such generalization is allowed and, therefore, true, see [3].

TABLE IV
BIN CAPACITY MAP

| Word Letters Left (WLL) | Pattern ...J | Pattern ...$K^1$ | Pattern ...$K^2$ | Pattern ...$K^3$ | Rem. |
|---|---|---|---|---|---|
| ⋮ | ⋮ | ⋮ | ⋮ | ⋮ | numbers UNDER the lines describe the bins ↓ for ↓ |
| | $^{(i+0)}c_0$ → + (add) | | | | |
| | $^{(i+1)}c_0$ → − = → | $^{(i+1)}c_1$ → + (add) | | | |
| | $^{(i+2)}c_0$ → | → − = → | $^{(i+2)}c_2$ | (copy) = → $^{(i+2)}c_3$ | |
| | $^{(i+3)}c_0$ → | → | → = | | |
| 17 of $L$ | ⋮ | ⋮ | ⋮ | ⋮ | ⋮ |
| 16 of $L$ | 19,079 | 17,697 | 15,033 | 9,898 | |
| 15 of $L$ | 9,898 | 9,181 | 7,799 | 5,135 | $L = 15$ 15,033 words total |
| 14 of $L$ | 5,135 | 4,763 | 4,046 | 2,664 | |
| 13 of $L$ | 2,664 | 2,471 | 2,099 | 1,382 | |
| 12 of $L$ | 1,382 | 1,282 | 1,089 | 717 | |
| 11 of $L$ | 717 | 665 | 565 | 372 | |
| 10 of $L$ | 372 | 345 | 293 | 193 | $L = 10$ 565 words total |
| 9 of $L$ | 193 | 179 | 152 | 100 | |
| 8 of $L$ | 100 | 93 | 79 | 52 | |
| 7 of $L$ | 52 | 48 | 41 | 27 | |
| 6 of $L$ | 27 | 25 | 21 | 14 | |
| 5 of $L$ | 14 | 13 | 11 | 7 | $L = 5$ 21 words total |
| 4 of $L$ | 7 | 7 | 6 | 4 | |
| 3 of $L$ | 4 | 3 | 3 | 2 | |
| 2 of $L$ | 2 | 2 | 1 | 1 | |
| 1 of $L$ | 1 | 1 | 1 | — | |
| 0 of $L$ (*implicit*) | — | undefined | undefined | undefined | no real bin exists further |
| $L$ to 1 | $L$ bins total | $L-1$ bins | $L-2$ bins | $L-3$ bins | |

TABLE V
BIN FREQUENCY MAP

| $L$ − WLL | Pattern ...J | Pattern ...$K^1$ | Pattern ...$K^2$ | Pattern ...$K^3$ | Rem. |
|---|---|---|---|---|---|
| $0 < L$ (*explicit*) | 1 | — | — | 1 | $L = 5$ 21 words total |
| $1 < L$ | 2 | 1 | — | — | |
| $2 < L$ | 3 | 2 | 1 | — | |
| $3 < L$ | 6 | 3 | 2 | 1 | |
| $4 < L$ | 12 | 6 | 3 | 2 | |
| $5 < L$ | 23 | 12 | 6 | 3 | $L = 10$ 565 words total |
| $6 < L$ | 44 | 23 | 12 | 6 | |
| $7 < L$ | 85 | 44 | 23 | 12 | |
| $8 < L$ | 164 | 85 | 44 | 23 | |
| $9 < L$ | 316 | 164 | 85 | 44 | |
| $10 < L$ | 609 | 316 | 164 | 85 | $L = 15$ 15,033 words total |
| $11 < L$ | 1,174 | 609 | 316 | 164 | |
| $12 < L$ | 2,263 | 1,174 | 609 | 316 | |
| $13 < L$ | 4,362 | 2,263 | 1,174 | 609 | |
| $14 < L$ | 8,408 | 4,362 | 2,263 | 1,174 | |
| $15 < L$ | 16,207 | 8,408 | 4,362 | 2,263 | $L = 20$ 400,025 words total |
| $16 < L$ | 31,240 | 16,207 | 8,408 | 4,362 | |
| $17 < L$ | ⋮ | ⋮ | ⋮ | ⋮ | ⋮ |
| ⋮ | $^{(i+0)}f_0$ → = (copy) | | | | |
| | = → | $^{(i+1)}f_1$ → = (copy) | | | |
| | | = → | $^{(i+2)}f_2$ → = (copy) | | |
| | | | = → | $^{(i+3)}f_3$ | numbers ABOVE the lines describe the bins ↑ for ↑ |
| $L-L$ to $L-1$ | $L$ bins total | $L-1$ bins | $L-2$ bins | $L-3$ bins | |

TABLE VI
BCM EVALUATION TIPS

| Natural Way—Regressive | Return Way—Progressive |
|---|---|
| Given: $L>0$, $H>0$, $\mathbf{C}$ = D-Rule ($H\times H$), $\mathbf{c}_s$ = BCM Seed ($H\times 1$), $0\le i<L$, $0\le r<H$. | |
| $^{(L-1)}\mathbf{c} = \mathbf{c}_s$ | $^{(0)}\mathbf{c} = \mathbf{C}^{L-1} \times \mathbf{c}_s$ |
| $^{(L-2)}\mathbf{c} = \mathbf{C} \times {}^{(L-1)}\mathbf{c} = \mathbf{C}^{1} \times \mathbf{c}_s$ | $^{(1)}\mathbf{c} = \mathbf{C}^{-1} \times {}^{(0)}\mathbf{c} = \mathbf{C}^{-1} \times {}^{(0)}\mathbf{c}$ |
| $^{(L-3)}\mathbf{c} = \mathbf{C} \times {}^{(L-2)}\mathbf{c} = \mathbf{C}^{2} \times \mathbf{c}_s$ | $^{(2)}\mathbf{c} = \mathbf{C}^{-1} \times {}^{(1)}\mathbf{c} = \mathbf{C}^{-2} \times {}^{(0)}\mathbf{c}$ |
| • • • • • • • • • • • • • • • • • • | • • • • • • • • • • • • • • • • • • |
| $^{(i)}\mathbf{c} = \mathbf{C} \times {}^{(i+1)}\mathbf{c} = \mathbf{C}^{L-i-1} \times \mathbf{c}_s$ | $^{(i)}\mathbf{c} = \mathbf{C}^{-1} \times {}^{(i-1)}\mathbf{c} = \mathbf{C}^{-i} \times {}^{(0)}\mathbf{c}$ |
| • • • • • • • • • • • • • • • • • • | • • • • • • • • • • • • • • • • • • |
| $^{(1)}\mathbf{c} = \mathbf{C} \times {}^{(2)}\mathbf{c} = \mathbf{C}^{L-2} \times \mathbf{c}_s$ | $^{(L-2)}\mathbf{c} = \mathbf{C}^{-1} \times {}^{(L-3)}\mathbf{c} = \mathbf{C}^{-L+2} \times {}^{(0)}\mathbf{c}$ |
| $^{(0)}\mathbf{c} = \mathbf{C} \times {}^{(1)}\mathbf{c} = \mathbf{C}^{L-1} \times \mathbf{c}_s$ | $^{(L-1)}\mathbf{c} = \mathbf{C}^{-1} \times {}^{(L-2)}\mathbf{c} = \mathbf{C}^{-L+1} \times {}^{(0)}\mathbf{c}$ |
| Taken: $^{(i)}\mathbf{c} = [\,^{(i)}c_0 \cdots {}^{(i)}c_{H-1}\,]^T$, where $^{(i)}c_r$ is kept set when $^{(i)}f_r > 0$ and reset otherwise. | |

TABLE VII
BFM EVALUATION TIPS

| Natural Way—Progressive | Return Way—Regressive |
|---|---|
| Given: $L>0$, $H>0$, $\mathbf{F}$ = U-Rule ($H\times H$), $\mathbf{f}_s$ = BFM Seed ($H\times 1$), $0\le i<L$, $0\le r<H$. | |
| $^{(0)}\mathbf{f} = \mathbf{f}_s$ | $^{(L-1)}\mathbf{f} = \mathbf{F}^{L-1} \times \mathbf{f}_s$ |
| $^{(1)}\mathbf{f} = \mathbf{F} \times {}^{(0)}\mathbf{f} = \mathbf{F}^{1} \times \mathbf{f}_s$ | $^{(L-2)}\mathbf{f} = \mathbf{F}^{-1} \times {}^{(L-1)}\mathbf{f} = \mathbf{F}^{-1} \times {}^{(L-1)}\mathbf{f}$ |
| $^{(2)}\mathbf{f} = \mathbf{F} \times {}^{(1)}\mathbf{f} = \mathbf{F}^{2} \times \mathbf{f}_s$ | $^{(L-3)}\mathbf{f} = \mathbf{F}^{-1} \times {}^{(L-2)}\mathbf{f} = \mathbf{F}^{-2} \times {}^{(L-1)}\mathbf{f}$ |
| • • • • • • • • • • • • • • • • • • | • • • • • • • • • • • • • • • • • • |
| $^{(i)}\mathbf{f} = \mathbf{F} \times {}^{(i-1)}\mathbf{f} = \mathbf{F}^{i} \times \mathbf{f}_s$ | $^{(i)}\mathbf{f} = \mathbf{F}^{-1} \times {}^{(i+1)}\mathbf{f} = \mathbf{F}^{-L+i+1} \times {}^{(L-1)}\mathbf{f}$ |
| • • • • • • • • • • • • • • • • • • | • • • • • • • • • • • • • • • • • • |
| $^{(L-2)}\mathbf{f} = \mathbf{F} \times {}^{(L-3)}\mathbf{f} = \mathbf{F}^{L-2} \times \mathbf{f}_s$ | $^{(1)}\mathbf{f} = \mathbf{F}^{-1} \times {}^{(2)}\mathbf{f} = \mathbf{F}^{-L+2} \times {}^{(L-1)}\mathbf{f}$ |
| $^{(L-1)}\mathbf{f} = \mathbf{F} \times {}^{(L-2)}\mathbf{f} = \mathbf{F}^{L-1} \times \mathbf{f}_s$ | $^{(0)}\mathbf{f} = \mathbf{F}^{-1} \times {}^{(1)}\mathbf{f} = \mathbf{F}^{-L+1} \times {}^{(L-1)}\mathbf{f}$ |
| Taken: $^{(i)}\mathbf{f} = [\,^{(i)}f_0 \cdots {}^{(i)}f_{H-1}\,]^T$, where $^{(i)}f_r$ is kept set when $^{(i)}c_r > 0$ and reset otherwise. | |

Assuming a couple of numbers reflecting how many times we repeat a so unified bin against how many word inidices it spans for, respectively, as a node, as well as clearly inking ties present in between such the nodes, we complete the definition of such a plot, see Table II and then Table I again.

Thence, we suppose we may record (describe) a framework of a code as more compactly as much compact we can set up (construct) an appropriate plot for such a code.

## RULE OF A CODE

Given a plot of a code, we describe a rule of that code with that plot, constructing such a rule the following way.

We place the nodes of a plot into (a set of) points of a two-dimensional rectangular grid with the axes indexed $0 \le i < L$ across $0 \le r < H$, where $H$ is the number of distinct syllable patterns, so the nodes of one given attribution, of letter period or syllable pattern, respectively, show the same index.

TABLE VIII
DESIGN RESOURCE DEMAND

| Parameter · · · · · measured | | Formula | $L = 5$ | $L = 8$ | $L = 10$ | $L = 15$ | $L = 16$ | $L = 20$ | $L = 24$ | $L = 25$ | $L = 30$ | $L = 32$ | $L = 35$ | $L = 40$ |
|---|---|---|---|---|---|---|---|---|---|---|---|---|---|---|
| Capacity, | in words | $\Sigma\, {}^{(0)}c_r = \Sigma\, {}^{(L-1)}f_r$ | 21 | 152 | 565 | ~15k | ~29k | ~400k | ~5.5M | ~11M | ~283M | ~1.1G | ~7.5G | ~201G |
| ALU bus width, | in bits | ceil $\log_2$ capacity | 5 | 8 | 10 | 14 | 15 | 19 | 23 | 24 | 29 | 30 | 33 | 38 |
| BCM ROM size, | in bits | $L \times$ ceil $\log_2 {}^{(0)}c_0$ | 20 | 56 | 90 | 210 | 240 | 380 | 528 | 575 | 840 | 960 | 1,155 | 1,480 |
| *w/o leading zeros,* | in bits | $\Sigma$ ceil $\log_2 {}^{(i)}c_0$ | 15 | 27 | 35 | 54 | 58 | 73 | 87 | 91 | 111 | 119 | 130 | 147 |
| BFM ROM size, | in bits | $L \times$ ceil $\log_2 {}^{(L-1)}f_0$ | 20 | 56 | 90 | 210 | 224 | 360 | 528 | 575 | 840 | 960 | 1,120 | 1,480 |
| *w/o leading zeros,* | in bits | $\Sigma$ ceil $\log_2 {}^{(i)}f_0$ | 10 | 22 | 30 | 50 | 53 | 66 | 82 | 86 | 106 | 114 | 125 | 142 |

TABLE IX
COMPACT $L = 10$ ENDEC FRAMEWORK EXAMPLES

| Max K's Run<br>leading [in letters]<br>inner/inter<br>trailing | BFM Seed | Afore<br>(worst case) | $t = 10m$ | $t+1$ | $t+2$ | $t+3$ | $t+4$ | $t+5$ | $t+6$ | $t+7$ | $t+8$ | $t+9$ | After<br>(worst case) | BCM Seed | Rem. |
|---|---|---|---|---|---|---|---|---|---|---|---|---|---|---|---|
| — 3 = 3<br>(gives 401 words) | $\begin{bmatrix}\blacksquare\\ \\ \\ \end{bmatrix}$ | •<br>•<br>•<br>$K^{\le 3}$ | 1 × 401 | 1 × 208<br>1 × 193 | 2 × 108<br>1 × 100<br>1 × 85 | 4 × 56<br>2 × 52<br>1 × 44<br>1 × 29 | 8 × 29<br>4 × 27<br>2 × 23<br>1 × 15 | 15 × 15<br>8 × 14<br>4 × 12<br>2 × 8 | 29 × 8<br>15 × 7<br>8 × 6<br>4 × 4 | 56 × 4<br>29 × 4<br>15 × 3<br>8 × 2 | 108 × 2<br>56 × 2<br>29 × 2<br>15 × 1 | 208 × 1<br>108 × 1<br>56 × 1<br>29 × 1 | J | $\begin{bmatrix}\blacksquare\\ \blacksquare\\ \blacksquare\\ \blacksquare\end{bmatrix}$ | options with a labeled pair behave in time like mirrors for each other |
| 1 = 3 − 2<br>(gives 565 words)<br>Base-23.77 max | $\begin{bmatrix}\blacksquare\\ \\ \\ \blacksquare\end{bmatrix}$ | •<br>•<br>$K^{\le 2}$ | 1 × 372<br>1 × 193 | 2 × 193<br>1 × 179 | 3 × 100<br>2 × 93<br>1 × 79 | 6 × 52<br>3 × 48<br>2 × 41<br>1 × 27 | 12 × 27<br>6 × 25<br>3 × 21<br>2 × 14 | 23 × 14<br>12 × 13<br>6 × 11<br>3 × 7 | 44 × 7<br>23 × 7<br>12 × 6<br>6 × 4 | 85 × 4<br>44 × 3<br>23 × 3<br>12 × 2 | 164 × 2<br>85 × 2<br>44 × 1<br>23 × 1 | 316 × 1<br>164 × 1<br>85 × 1 | •<br>•<br>•<br>$K^{\le 3}$ | $\begin{bmatrix}\blacksquare\\ \blacksquare\\ \blacksquare\\ \end{bmatrix}$ | |
| 2 = 3 − 1<br>(gives 565 words)<br>Base-23.77 max | $\begin{bmatrix}\blacksquare\\ \\ \blacksquare\\ \end{bmatrix}$ | •<br>$K^{\le 1}$ | 1 × 316<br>1 × 249 | 2 × 164<br>1 × 152<br>1 × 85 | 4 × 85<br>2 × 79<br>1 × 67 | 7 × 44<br>4 × 41<br>2 × 35<br>1 × 23 | 14 × 23<br>7 × 21<br>4 × 18<br>2 × 12 | 27 × 12<br>14 × 11<br>7 × 9<br>4 × 6 | 52 × 6<br>27 × 6<br>14 × 5<br>7 × 3 | 100 × 3<br>52 × 3<br>27 × 3<br>14 × 2 | 193 × 2<br>100 × 1<br>52 × 1<br>27 × 1 | 372 × 1<br>193 × 1 | •<br>•<br>$K^{\le 2}$ | $\begin{bmatrix}\blacksquare\\ \blacksquare\\ \\ \end{bmatrix}$ | |
| 3 = 3 —<br>(gives 401 words)<br>Base-20.02 | $\begin{bmatrix}\blacksquare\\ \blacksquare\\ \\ \end{bmatrix}$ | J | 1 × 208<br>1 × 193 | 2 × 108<br>1 × 100<br>1 × 85 | 4 × 56<br>2 × 52<br>1 × 44<br>1 × 29 | 8 × 29<br>4 × 27<br>2 × 23<br>1 × 15 | 15 × 15<br>8 × 14<br>4 × 12<br>2 × 8 | 29 × 8<br>15 × 7<br>8 × 6<br>4 × 4 | 56 × 4<br>29 × 4<br>15 × 3<br>8 × 2 | 108 × 2<br>56 × 2<br>29 × 2<br>15 × 1 | 208 × 1<br>108 × 1<br>56 × 1<br>29 × 1 | 401 × 1 | •<br>$K^{\le 1}$ | $\begin{bmatrix}\blacksquare\\ \\ \\ \end{bmatrix}$ | |

TABLE X
FRAMEWORK IMPLEMENTATION GUIDELINE

| Stream Processing Behavior Required | Suitable Frameworks | | Bin Passing Order | Time Application Flow Assumed | Map Time Rel. | ALU Operations Needed |
|---|---|---|---|---|---|---|
| On the fly, letter by letter, when $t$ steps | $^{(i)}$**c**'s<br>$^{(i)}$**f**'s | of target BCM<br>of mirror BFM | with $i$ ascending<br>with $i$ descending | as is<br>$t^* = (t \operatorname{div} L) + (L - 1 - t \bmod L)$ | $i \equiv t \bmod L$<br>$i \equiv t^* \bmod L$ | addition<br>subtraction<br>comparison<br>all unsigned |
| On ready, by bulks of $L$ letters, delayed | $^{(i)}$**f**'s<br>$^{(i)}$**c**'s | of suitable BFM<br>of suitable BCM | with $i$ advancing | any from the two above | any linear | |

Having this done, see Table I yet again, we reflect the plot inter-node ties, directed as well as overlapped, obtaining the direction-related rule in a matrix form, see Table III.

Such an upward rule, i.e., applied along time ($t$) ascending, predetermines the content of a bin frequency map,[3] set BFM in short, related to such a code, see Tables V and VII.

Such a downward rule, i.e., applied along time descending, predetermines the content of a bin capacity map,[4] set BCM in short, related to such a code, see Tables IV and VI.

Based on the systematic nature of such maps, see Tables V and IV again, we can set up (construct) the respective part of a plot of such a code very compact, see Table VIII.

Having this one understood, we reflect the plot edge nodes, just alone as well as in the respective direction, obtaining the direction-related seed in a vector form, see Table IX.

Such an upward seed, i.e., applied once time ($t$) ascending under the respective rule, predetermines the content of a BFM, initializing its first vector by itself,[5] see Table VII again.

Such a downward seed, i.e., applied once time descending under the respective rule, predetermines the content of a BCM, initializing its last vector by itself,[6] see Table VI again.

Based on the systematic nature of such maps then, thus, of such plots, we can set up (construct) an appropriate framework of such a code very compact, too, see Table X.

## SEED OF A CODE

Given a rule of a plot of a code, we describe a seed of that code with that plot and that rule, to complete the definition of that code, constructing such a seed the following way.

Although a rule of a plot and its complement, i.e., opposite by the direction, rule of the same plot are mutually predefined as they are transposable into each other, see Table III again, a seed of a plot and its counterpart of the same plot, respectively, are just mutually bounded as restricting on each other.

Anyway, every one among the rules and the seeds of a plot of a code inherits from then responds for the implementation of the principles the code is designed with.

## GOAL OF A CODE

Given a task to design a code, we often get into a situation where many options fulfill the principles the code is designed with, originally, so, to make our choice as much as reasonable, we need to engage an extra measure, i.e., a goal.

Such a goal, being applied, updates the principles the code is designed with, that further results in a renewed framework, including its plot, then rules and seeds, and then bin maps, all corresponding to the updated principles, see Table XI.

Such a goal, being applied, may modify one, some, or even all of the properties of the code we design, as for its purpose, side effect, or both, useful or neutral, see Table XII.

[3]In the scope of a plot of a code, an element of its BFM, ${}^{(i)}\mathrm{f}_r$, indicates how many repetitions of a unified bin are within the respective node.

[4]In the scope of a plot of a code, an element of its BCM, ${}^{(i)}\mathrm{c}_r$, indicates how many word indices are in the unified bin of the respective node.

[5]In the scope of a plot of a code, a unity set in the first vector of its BFM, i.e., in ${}^{(0)}\mathbf{f}$, enables a word of the code to start at the respective node.

[6]In the scope of a plot of a code, a unity set in the last vector of its BCM, i.e., in ${}^{(L-1)}\mathbf{c}$, enables a word of the code to end at the respective node.

TABLE XI
COMPACT $L = 15$ ENDEC FRAMEWORK EXAMPLES

Goal: Prevent both K's and J's runs of >3 letters

U-Rule = **F** = **C**ᵀ = D-Ruleᵀ — enables 2 • (2 + 2 + 1) = 2 • 5 = 10 transits in a (3 + 3) x (3 + 3) = 6 x 6 transit space (column = from, row = into):

| row = into ← / column = from ↓ | J³ | J² | J¹ | K¹ | K² | K³ |
|---|---|---|---|---|---|---|
| J³ | | ■ | | | | |
| J² | | | ■ | | | |
| J¹ | | | | ■ | ■ | ■ |
| K¹ | ■ | ■ | ■ | | | |
| K² | | | | ■ | | |
| K³ | | | | | ■ | |

| Max K's Run (in letters) leading / inner/inter / trailing | Max J's Run (in letters) leading / inner/inter / trailing | BFM Seed = $^{(t+0)}\mathbf{f}$ defines BFM's first 3 + 3 = 6 bins (J³ J² J¹ K¹ K² K³) | BCM Seed = $^{(t+14)}\mathbf{c}$ defines BCM's last 3 + 3 = 6 bins (J³ J² J¹ K¹ K² K³) | $^{(t+0)}\mathbf{f}^{T} \bullet {}^{(t+0)}\mathbf{c} = \ldots = {}^{(t+i)}\mathbf{f}^{T} \bullet {}^{(t+i)}\mathbf{c} = \ldots = {}^{(t+14)}\mathbf{f}^{T} \bullet {}^{(t+14)}\mathbf{c}$ Most Binary Spaces within... | Capacity | Eq. Base |
|---|---|---|---|---|---|---|
| — 3 = 3 | — 3 = 3 | [ empty ]ᵀ | [ empty ]ᵀ | this configuration results in no words | EMPTY | — |
| | 1 = 3 − 2 | [■ · · · · ·]ᵀ | [· ■ ■ ■ ■ ■]ᵀ | 4 + 64 + 256 + 512 + 2,048 = | 2,884 | 14.23 |
| | 2 = 3 − 1 | [· ■ · · · ·]ᵀ | [· · ■ ■ ■ ■]ᵀ | ... + 512 + 1,024 + 2,048 < | 3,737 | 15.52 |
| | 3 = 3 — | [· · ■ · · ·]ᵀ | [· · · ■ ■ ■]ᵀ | 4 + 64 + 256 + 512 + 2,048 = | 2,884 | 14.23 |
| 1 = 3 − 2 | — 3 = 3 | [· · · · · ■]ᵀ | [■ ■ ■ ■ ■ ·]ᵀ | 4 + 64 + 256 + 512 + 2,048 = | 2,884 | 14.23 |
| | 1 = 3 − 2 | [■ · · · · ■]ᵀ | [· ■ ■ ■ ■ ·]ᵀ | 16 + 128 + 1,024 + 4,096 = | 5,264 | 17.40 |
| | 2 = 3 − 1 | [· ■ · · · ■]ᵀ | [· · ■ ■ ■ ·]ᵀ | ... + 256 + 1,024 + 4,096 < | 5,516 | 17.67 |
| | 3 = 3 — | [· · ■ · · ■]ᵀ | [· · · ■ ■ ·]ᵀ | ... + 512 + 1,024 + 2,048 < | 3,737 | 15.52 |

TABLE XII
ALIGNMENT RELATED PROPERTIES

| Max K's Run (leading inner trailing) | Max J's Run (leading inner trailing) | JUMP Probability of Occurrence, $p$(J), at the $i$-th Letter Period in the Loop: ••• $L-4$ | $L-3$ | $L-2$ | $L-1=i=$ (comma) | 0 | 1 | 2 | 3 ••• | Kind (of $p$) | Comma Ambit Shape (coarse, not for scale) | Eq. Base = Capacity$^{5/L}$: $L=20$ | $L=30$ | $L=40$ |
|---|---|---|---|---|---|---|---|---|---|---|---|---|---|---|
| 1 = 3 − 2 | not limited | .5 | .6 | .6 | .6 | .7 | .7 | .5 | .6 | approx. | | 25.15 | 25.63 | 25.87 |
| 1 = 3 − 2 | — 3 — | .5 | .6 | .6 | — | — | 1. | .5 | .5 | approx. | | 12.90 | 15.19 | 16.48 |
| | — 3 > 1 | .5 | .6 | .4 | .4 | — | 1. | .5 | .5 | approx. | | 14.62 | 16.51 | 17.54 |
| | — 3 > 2 | .5 | .5 | .5 | .5 | — | 1. | .5 | .5 | approx. | | 15.34 | 17.05 | 17.97 |
| | — 3 = 3 | .5 | .5 | .5 | .5 | — | 1. | .5 | .5 | approx. | | 15.70 | 17.31 | 18.18 |
| | 1 = 3 − 2 | .5 | .5 | .5 | .5 | .5 | .5 | .5 | .5 | exact | | 18.24 | 19.14 | 19.60 |
| | 2 = 3 − 1 | .5 | .6 | .4 | .4 | .6 | .6 | .4 | .5 | approx. | | 18.46 | 19.29 | 19.71 |
| | 3 = 3 — | .5 | .6 | .6 | — | .6 | .6 | .5 | .5 | approx. | | 16.75 | 18.07 | 18.78 |
| | 2 < 3 — | .5 | .6 | .6 | — | .6 | .6 | .4 | .5 | approx. | | 16.29 | 17.75 | 18.52 |
| | 1 < 3 — | .5 | .6 | .6 | — | .5 | .5 | .5 | .5 | approx. | | 15.34 | 17.05 | 17.97 |
| 1 < 3 − 1 | not limited | .6 | .5 | .7 | .7 | .7 | .7 | .5 | .6 | approx. | | 24.14 | 24.94 | 25.35 |
| 1 < 3 − 1 | 2 = 3 − 1 | .5 | .5 | .5 | .5 | .6 | .6 | .4 | .5 | approx. | | 17.38 | 18.53 | 19.13 |
| | 3 = 3 — | .5 | .5 | 1. | — | .6 | .6 | .5 | .5 | approx. | | 15.02 | 16.81 | 17.78 |
| | 2 < 3 — | .5 | .5 | 1. | — | .6 | .6 | .4 | .5 | approx. | | 14.62 | 16.51 | 17.54 |
| | 1 < 3 — | .5 | .5 | 1. | — | .5 | .5 | .5 | .5 | approx. | | 13.76 | 15.86 | 17.02 |

Such a goal, being applied, results, finally, in a new coding means based on a plot set up so it reaches that goal.[7]

## CONCLUSION

As we can notice, constructing a plot (of a framework) of a code helps much in discovering a systematic portion of such (a framework of) a code, the portion translatable—via analytic measures, various but rational—into a smaller description, that constitutes the design way considered in this paper.

On this way, we try to find out a code whose framework is representable by a compact plot,[8] that makes such a framework as well as a coding means based on it also compact, that next, in its turn now, enables for us, when we employ such a means, to say we generate quasi base-21 words compactly.

[7]Generally speaking, a plot is a comprehensive expression of the respective framework of a code we consider in the paper, therefore, when such is allowed and true, we may equate between such a plot and such a framework, arbitrarily substituting each of the terms with each other in our consideration.

Speaking further, a bin map is an expression of the respective plot and thus of the respective framework, sufficient to implement a coding means capable to deal with the respective code, see Table X, therefore, when such is allowed and true, we may equate between, too, while losing nothing sensible.

[8]Reading plot, we do it like it is in a role-playing game, or in an interactive TV show, where the plot describes all the permissible among all the possible scenarios the player, or the subscriber, can travel from (one of) the beginning, into (one of) the ending, through (a chain of) intermediate scenes, in all which such the client manifests initially, then occasionally, then finally, respectively, the choice of that precise route the client desires to travel along, of every time the client is involved in. Therefore, based on such, we could label the shown
in Table I: // as basic as is
an EBTO plot code, that is a $L(5)$ $N(21)$ *QBTO*$(21.00)$ plot code ;
in Table IX: // limit on a K's run
a $L(10)$ $N(401)$ $\mathtt{K}(0,3,3)$ *QBTO*$(20.02)$ plot code ; // $\emptyset$T···T$o$
a $L(10)$ $N(565)$ $\mathtt{K}(1,3,2)$ *QBTO*$(23.77)$ plot code ; // $x$T···T$y$
a $L(10)$ $N(565)$ $\mathtt{K}(2,3,1)$ *QBTO*$(23.77)$ plot code ; // $y$T···T$x$
a $L(10)$ $N(401)$ $\mathtt{K}(3,3,0)$ *QBTO*$(20.02)$ plot code ; // $o$T···T$\emptyset$
in Table XI: // limits on K's and J's runs
a $L(15)$ $N(2{,}884)$ $\mathtt{K}(0,3,3)$ $\mathtt{J}(1,3,2)$ *QBTO*$(14.23)$ plot code ;
..................................................................
a $L(15)$ $N(3{,}737)$ $\mathtt{K}(1,3,2)$ $\mathtt{J}(3,3,0)$ *QBTO*$(15.52)$ plot code ;
in Table XII: // limits on K's and J's runs
a $L(20)$ $N(2^{14.76})$ $\mathtt{K}(1,3,2)$ $\mathtt{J}(0,3,0)$ *QBTO*$(12.90)$ plot code ;
..................................................................
a $L(40)$ $N(2^{32.71})$ $\mathtt{K}(1,3,1)$ $\mathtt{J}(1,3,0)$ *QBTO*$(17.02)$ plot code ;
from top to bottom then from left to right, if any, as mentioned in the tables, respectively, excepting the first row of the last two tables, which are omitted.

## REFERENCES

[1] A. Ivanov, "Base-21 scrambling," .................................. ......... @*arXiv*, doi:10.48550/arXiv.yymm.nnnn (bundle), pp. 01–04.
[2] A. Ivanov, "Base-21 word alignment and boundary detection," ........ ......... @*arXiv*, doi:10.48550/arXiv.yymm.nnnn (bundle), pp. 05–08.
[3] A. Ivanov, "Quasi base-21 words," ................................. ......... @*arXiv*, doi:10.48550/arXiv.yymm.nnnn (bundle), pp. 09–12.

# Quasi Base-21 Words Balanced on the Framework

Alexander Ivanov

***Abstract*—In this paper, we further develop on the category of codes collectively called quasi base-21 words, or QBTO in short, derived from exact base-21 words, or EBTO in short, that are a subset as well as the root of the offspring, now focusing on their balancing problems and exercises.**

## INTRODUCTION

BASE-21 words, including the so called progenitor,[1] then exact,[2] and then quasi,[3] as considered in [1] and [2] then in [3] and [4], respectively, are manageable codes intended to improve on the originally given properties of the line code of a coding means we need to design, fix, or upgrade.

The manageability of such a code refers to the probabilistic properties of the code, which altogether define the balance of the code, and is based completely on the underlying structure of such a code, representable by a framework [3].

By its turn, a framework of such a code is implementable—in an appropriate coding means—many ways sourcing out of its plot, a very useful among those is a bin map [4].

In this paper, we plan to balance such a code, manipulating on its BPM,[4] see Tables I and II, as the uniform reflection of its framework capable to sufficiently describe a framework of such a code, in its maternal,[5] delta, and balanced states, both individually and coherently, see Tables III and IV.[6]

Recalling the fate of submission of many prior works to the peer reviewed journal, such a try with this one also promises no chance, probably.

Please sorry for the author has no time to find this work a new home, peer reviewed or not, except of arXiv, and just hopes there it meets its reader, one or maybe various, whom the author beforehand thanks for their regard.

A. Ivanov is with JSC Continuum, Yaroslavl, the Russian Federation.

[1]The base-21 words are the progenitor of both next exact and quasi base-21 words as well as a valid example of both exact and quasi base-21 codes, that defines a set of $3\times7=21$ five-letter-long distinct serial images patterned $x\mathrm{T}y$, where $x \in \{1,2,3\}$ and $y \in \{1,2,3,4,5,6,7\}$, see [1] and [2].

[2]Exact base-21 words, or EBTO codes, are of those whose equivalent base is exact 21, where the base is a comparative measure referenced to that value of the progenitor. We consider an exact base-21 code as a valid quasi base-21 code, assuming such generalization is allowed and true, see [3].

[3]Quasi base-21 words, or QBTO codes, are of those whose equivalent base is about 21, near or far, as well as whose structure is clearly representable by a framework, see [3]. We consider any appropriate accessory, e.g., a bin map, frequency-related or capacity-related, as a valid, compact(-ized) expression of and, thus, a valid, compact substitute for such a framework of a quasi base-21 code, assuming such generalization is allowed and true, too, see [4].

[4]P in this abbreviation may stand for product, production, portion, partition, part, i.e., any suitable term for a contribution into the whole quantity.

[5]The maternal state of (a framework of a) a code is the state corresponding to the time when we set up (construct) the respective plot of that code.

[6]In this paper, we manipulate on the code denoting for the base-21 words, because it is enough short but vivid to be illustrative in our consideration.

TABLE I
BIN PRODUCT MAP

| Bin Frequency Map | Bin Product Map | Bin Capacity Map |
|---|---|---|
| $\overrightarrow{r}$, $\downarrow i$: · · · $^{(i)}\mathrm{f}_r$ · · · | $\overrightarrow{r}$, $\downarrow i$: · · · $^{(i)}\mathrm{p}_r$ · · · | $\overrightarrow{r}$, $\downarrow i$: · · · $^{(i)}\mathrm{c}_r$ · · · |
| $^{(i)}\mathrm{p}_r = {}^{(i)}\mathrm{f}_r \times {}^{(i)}\mathrm{c}_r$ when both $^{(i)}\mathrm{f}_r$ and $^{(i)}\mathrm{c}_r$ exist, zero otherwise; $i = *, 0 \ldots L-1$; $r = 0 \ldots H-1$ | | |

TABLE II
$L = 5$ RESPECTIVE MAPS

| $i$ | BFM | | | | $\Sigma_r\,{}^{(i)}\mathrm{f}_r$ | BPM | | | | $\Sigma_r\,{}^{(i)}\mathrm{p}_r$ | BCM | | | | $\Sigma_r\,{}^{(i)}\mathrm{c}_r$ |
|---|---|---|---|---|---|---|---|---|---|---|---|---|---|---|---|
| * (implicit) | — | — | 1 | — | 1 | — | — | 21 | — | 21 | — | — | 21 | — | 21 |
| 0 | 1 | — | — | 1 | 2 | 14 | — | — | 7 | 21 | 14 | — | — | 7 | 21 |
| 1 | 2 | 1 | — | — | 3 | 14 | 7 | — | — | 21 | 7 | 7 | — | — | 14 |
| 2 | 3 | 2 | 1 | — | 6 | 12 | 6 | 3 | — | 21 | 4 | 3 | 3 | — | 10 |
| 3 | 6 | 3 | 2 | 1 | 12 | 12 | 6 | 2 | 1 | 21 | 2 | 2 | 1 | 1 | 6 |
| 4 | 12 | 6 | 3 | — | 21 | 12 | 6 | 3 | — | 21 | 1 | 1 | 1 | — | 3 |

TABLE III
BPM-BASED BALANCING

| Maternal BPM | | Delta BPM | | Balanced BPM |
|---|---|---|---|---|
| $\begin{pmatrix} \cdot & \cdot & \cdot & \cdot & \cdot \\ \cdot & \cdot & ^{(i)}\mathrm{p}_r & \cdot & \cdot \\ \cdot & \cdot & \cdot & \cdot & \cdot \end{pmatrix}$ | − | $\begin{pmatrix} \cdot & \cdot & \cdot & \cdot & \cdot \\ \cdot & \cdot & ^{(i)}\Delta\mathrm{p}_r & \cdot & \cdot \\ \cdot & \cdot & \cdot & \cdot & \cdot \end{pmatrix}$ | = | $\begin{pmatrix} \cdot & \cdot & \cdot & \cdot & \cdot \\ \cdot & & ^{(i)}\mathrm{p}_r - {}^{(i)}\Delta\mathrm{p}_r & & \cdot \\ \cdot & \cdot & \cdot & \cdot & \cdot \end{pmatrix}$ |
| Base-$(N_\mathrm{M})^{5/L}$, $N_\mathrm{M} = N_\mathrm{M}(L)$ | | $N_\mathrm{M} \longrightarrow N_\mathrm{B} : 0 \le {}^{(i)}\Delta\mathrm{p}_r \le {}^{(i)}\mathrm{p}_r$ | | Base-$(N_\mathrm{B})^{5/L}$, $N_\mathrm{B} = N_\mathrm{M} - N_\Delta$ |

TABLE IV
EXAMPLE BASE-21 TO BASE-16 BALANCING

| $i$ | $p_\mathrm{M}(\mathrm{J})$ | M-BPM | | | | Δ-BPM | | | | B-BPM | | | | $p_\mathrm{B}(\mathrm{J})$ |
|---|---|---|---|---|---|---|---|---|---|---|---|---|---|---|
| * | n/a | — | — | 21 | — | — | — | 5 | — | — | — | 16 | — | n/a |
| 0 | .67 | 14 | — | — | 7 | 3 | — | — | 2 | 11 | — | — | 5 | $.69^{+.02}$ |
| 1 | .67 | 14 | 7 | — | — | 3 | 2 | — | — | 11 | 5 | — | — | $.69^{+.02}$ |
| 2 | .57 | 12 | 6 | 3 | — | 4 | 1 | — | — | 8 | 5 | 3 | — | $.50^{-.07}$ |
| 3 | .57 | 12 | 6 | 2 | 1 | 4 | — | 1 | — | 8 | 6 | 1 | 1 | $.50^{-.07}$ |
| 4 | .57 | 12 | 6 | 3 | — | 4 | 1 | — | — | 8 | 5 | 3 | — | $.50^{-.07}$ |
| $\max\{\Delta p\} = .10$ | | Base-21 | | | | Base-21 → Base-16 | | | | Base-16 | | | | $.19^{+.09}$ |

## DECOMPOSITION OF A CODE

Given such a code, whose length is $L$ while capacity is $N$, we recognize its BPM as a superposition of distinct BPMs of the distinct words the code denotes for, see Table V.[7]

[7]In this paper, we consider up to $L+1$ letter periods of a map, beginning either from $* = 0{-}1$, i.e., $i \geq *$, or from $0 = 0$, i.e., $i \geq 0$, purposefully.

TABLE V
BPM DECOMPOSITION RULE

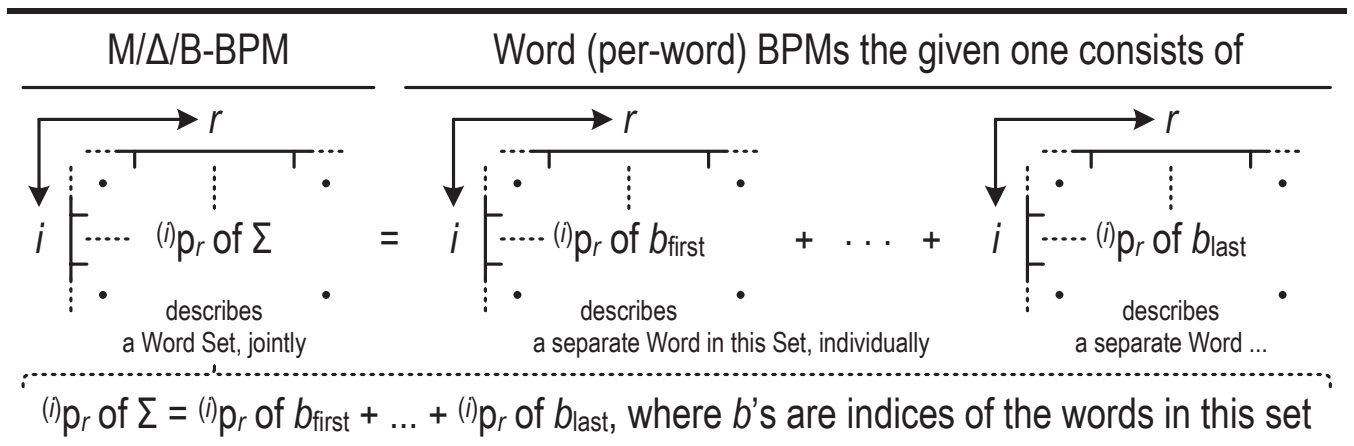

TABLE VI
BPM CONSTRUCTION MEMO

| Parameter | M/Δ/B-BPM | Word BPM |
|---|---|---|
| Map originality, for the same capacity choice made | unique or shared | unique among all $b$'s |
| Map element—possible values | $0 \le {}^{(i)}p_r \le N_M$ | $0 \le {}^{(i)}p_r \le 1$ (▪) |
| Map element—considered as | ${}^{(i)}p_r$ of Σ | ${}^{(i)}p_r$ of (the given) $b$ |
| Map as a whole—key properties | $\Sigma_r {}^{(i)}p_r$ = capacity<br>$0 <$ capacity $\le N_M$ | $\Sigma_r {}^{(i)}p_r = 1$<br>$\Sigma_{i(\neq *)} \Sigma_r {}^{(i)}p_r = L$ |

$N_M$ is the capacity of the ENDEC framework described by the (respective) M-BPM

TABLE VII
$L = 5$ WORD BPMS

| Image | JJJJJ | JJJJK | JJJKJ | JJJKK | JJKJJ | JJKJK | JJKKJ | JKJJJ | JKJJK | JKJKJ | JKJKK | JKKJJ | JKKJK | JKKKJ | KJJJJ | KJJJK | KJJKJ | KJJKK | KJKJJ | KJKJK | KJKKJ |
|---|---|---|---|---|---|---|---|---|---|---|---|---|---|---|---|---|---|---|---|---|---|
| Index | 0 | 1 | 2 | 3 | 4 | 5 | 6 | 7 | 8 | 9 | 10 | 11 | 12 | 13 | 14 | 15 | 16 | 17 | 18 | 19 | 20 |
| Alias | 3T7 | 3T6 | 3T5 | 3T4 | 3T3 | 3T2 | 3T1 | 2T7 | 2T6 | 2T5 | 2T4 | 2T3 | 2T2 | 2T1 | 1T7 | 1T6 | 1T5 | 1T4 | 1T3 | 1T2 | 1T1 |

TABLE VIII
BASE-21 ENCODE PROCESS

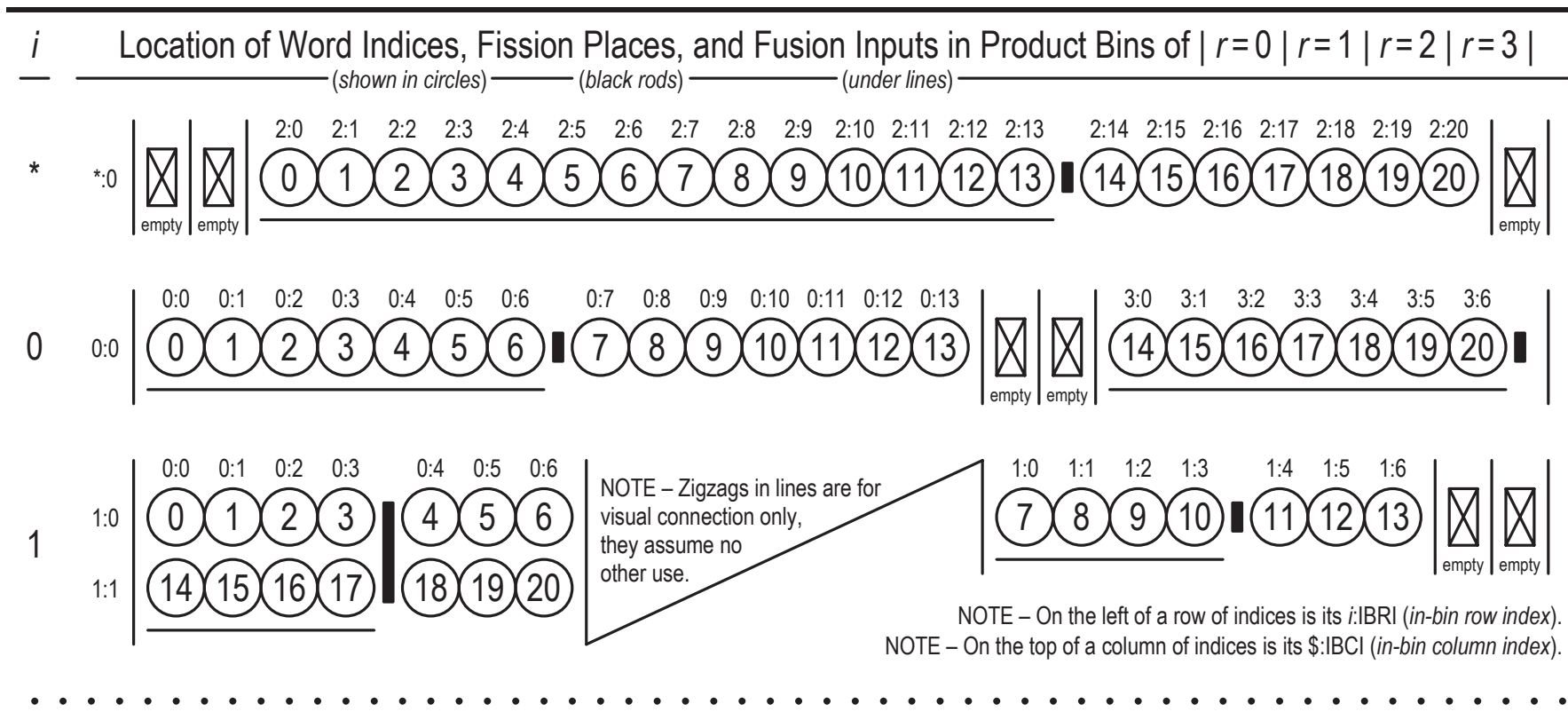

Index Location Algorithm, for a given $0 \le b < N_M$

INIT:
- $i \leftarrow *$, assuming below that $* = -1$ ; ${}^{(*)}$IBRI $\leftarrow 0$ ;
- ${}^{(*)}r \leftarrow r$ of the single bin of ${}^{(*)}c_r > 0$ ; ${}^{(*)}$IBCI $\leftarrow b$ .

Do the next $L$ times, aliasing $\$ \equiv {}^{(i)}r$, iterating $* \le i < L-1$ :

FISSION:
- ${}^{(i)}[B_B]_\$ \leftarrow {}^{(i)}c_\$$ ; $( {}^{(i)}[B_Z]_\$ \leftarrow 0 )$ ;
- ${}^{(i)}[B_F]_\$ \leftarrow$ ${}^{(i)}[B_F]_\$ \leftarrow {}^{(i)}[B_B]_\$ - {}^{(i+1)}c_{\$+1}$ if $\$ < H-1$ and ${}^{(i+1)}c_{\$+1} > 0$; ${}^{(i)}[B_B]_\$$ otherwise ;

FUSION:

| | if ${}^{(i)}$IBCI $< {}^{(i)}[B_F]_\$$ | otherwise | |
|---|---|---|---|
| ${}^{(i+1)}$IBRI $\leftarrow$ | ${}^{(i)}$IBRI $+ \Sigma_{r<\$} {}^{(i)}f_r$ | ${}^{(i)}$IBRI | ; |
| ${}^{(i+1)}r \leftarrow$ | 0 | $\$ + 1$ | ; |
| ${}^{(i+1)}$IBCI $\leftarrow$ | ${}^{(i)}$IBCI | ${}^{(i)}$IBCI $- {}^{(i)}[B_F]_\$$ | . |

WHERE: c's and f's are pre-known or recovered from p's of the M-BPM describing the respective ENDEC framework. (see BFM/BCM recovery from M-BPM)

TABLE IX
INDEX REJECTION MODEL

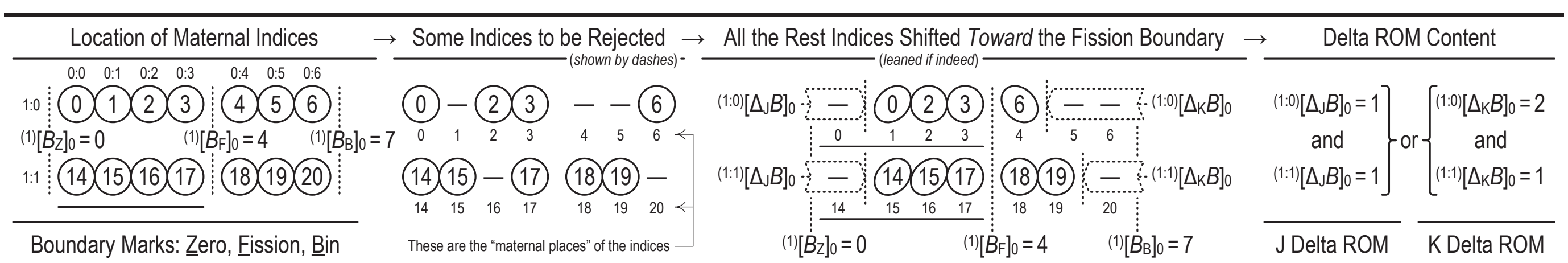

TABLE X
REJECT OPERATION RESOURCE DEMAND

| Parameter · · · · · measured | | Formula | L = 5 | L = 8 | L = 10 | L = 15 | L = 16 | L = 20 | L = 24 | L = 25 | L = 30 | L = 32 | L = 35 | L = 40 |
|---|---|---|---|---|---|---|---|---|---|---|---|---|---|---|
| Capacity limit, | in words | $N_M = N_M(L)$ | 21 | 152 | 565 | ~15k | ~29k | ~400k | ~5.5M | ~10.6M | ~283M | ~1.1G | ~7.5G | ~201G |
| ALU bus width limit, | in bits | ceil $\log_2 N_M$ | 5 | 8 | 10 | 14 | 15 | 19 | 23 | 24 | 29 | 30 | 33 | 38 |
| Delta ROM size limit, | in bits | (☼) using path a | 33 | 275 | 1,046 | ~28k | ~54k | ~749k | ~10.3M | ~19.9M | ~530M | ~2.0G | ~14.1G | ~375G |
| *w/o leading zeros,* | in bits | (☼) using path b | 29 | 256 | 973 | ~26k | ~50k | ~697k | ~9.6M | ~18.5M | ~493M | ~1.8G | ~13.1G | ~349G |
| *reduction percentage* | | $i = *,0...L-2$ ; $r = 0...H-1$ | −12.12% | −6.91% | −6.98% | −6.96% | −6.97% | −6.97% | −6.97% | −6.97% | −6.97% | −6.97% | −6.97% | −6.97% |

(☼) a: $\Sigma_i ( \max\{{}^{(i)}\beta_r\} \times \Sigma_r {}^{(i)}\lambda_r )$, b: $\Sigma_i \Sigma_r ( {}^{(i)}\beta_r \times {}^{(i)}\lambda_r )$, where ${}^{(i)}\beta_r = \text{ceil} \log_2 ( \min \{ {}^{(i+1)}c_0 , {}^{(i+1)}c_{r+1} \} + {}^{(i)}\varphi_r )$ if ${}^{(i+1)}c_0$ and ${}^{(i+1)}c_{r+1}$ exist, else 0; ${}^{(i)}\varphi_r = 1$ if ${}^{(i)}f_r > 1$, else 0; ${}^{(i)}\lambda_r = {}^{(i)}f_r$ if ${}^{(i)}\beta_r > 0$, else 0.

TABLE XI
EXAMPLE BASE-21 TO BASE-16 ENCODING

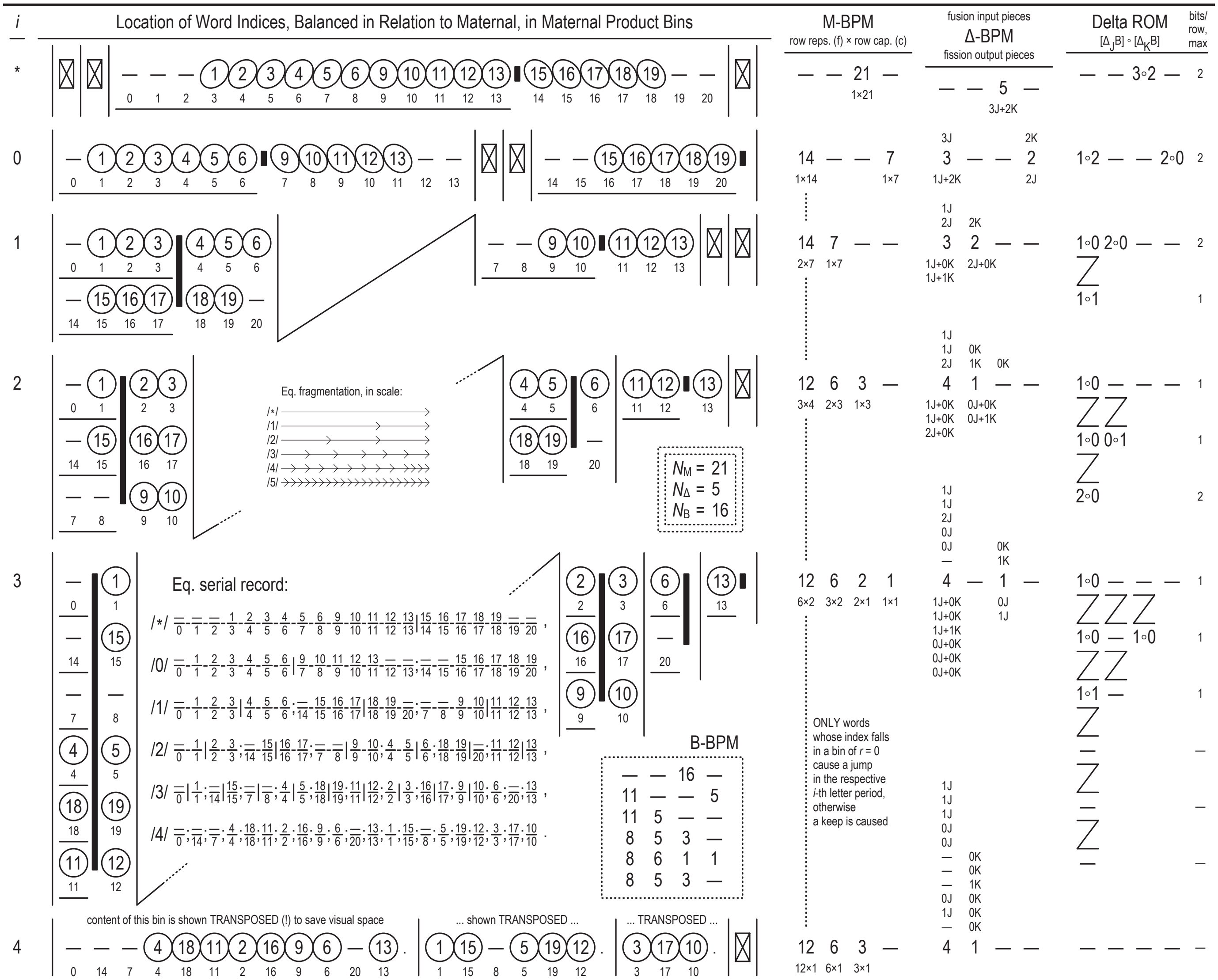

In the scope of such a code, there are only three such maps related with its states, including maternal (M),[8] delta ($\Delta$), and balanced (B), all of $N \geq 1$ and each of $N = N_{\mathrm{M}}$, $N = N_{\Delta}$, and $N = N_{\mathrm{B}}$, respectively, see Table VI, and exact $N_{\mathrm{M}}$ such maps related with (the distinct serial images of) its words, all of $N = N_0 = \cdots = N_{N_{\mathrm{M}}-1} = 1$, see Table VII.

Assigning a distinct (continuous) index to each word BPM, $0 \leq b < N_{\mathrm{M}}$, we establish an appropriate ENDEC process, as a fission–fusion procedure, during which such an index takes its unique place at the rectangular grid (in the space) of each of the respective $L + 1$ product bins, see Table VIII.

Thus, such a word in such a code features a distinct index, a distinct image, a distinct BPM, and a distinct chain of places in product bins of such a process, that sets up the ground for our further steps, see Tables V, VI, VII, and VIII again.

[8]A maternal map is self-sufficient, i.e., M-BPM ≡ framework, while all the rest are not, therefore any mention of a delta, balanced, or word map assumes an implicit reference to the respective maternal one and from that one further to the respective ENDEC framework, always as well as anyway.

## MODIFICATION OF A CODE

Given such a decomposed code, i.e., a code all whose maps are known, we balance the code via rejection,[9] see Table IX, of $N_{\Delta} = N_{\mathrm{M}} - N_{\mathrm{B}}$ out of $N_{\mathrm{M}}$ words it denotes for.

Such a rejection costs us a memory, see Table X, we should equip an appropriate coding means running such an ENDEC process with, the volume of that rises rapidly and enormously along the length of words to be rejected, i.e., along $L$.

However, a particular case may necessitate for a very much lesser volume of such a memory, because it may be in no need to store BPM-related information about every product bin, but only about those who are essential so to ensure the respective ENDEC process, as its underlying fission–fusion procedure, is run properly and unambiguously,[10] see Table XI.

[9]We read *exclusion* and *rejection* (reject operation) different, as mentioned in [3] then [4] and in this paper, respectively, supposing the former results in a new framework while the latter does not. Of course, both those operations may be involved to balance a code, making their effects aggregated.

[10]In the scope of the output of such a routine run by a coding means.

TABLE XII
INVERSE ENCODE PROCESS

| $i$ | Location of Word Indices Shifted *Fromward* the Fission Boundary of a Bin, in Product Bins | M-BPM | Δ-BPM | Inverse Δ-ROM $[B_F-\Delta_J B] \circ [B_B-B_F-\Delta_K B]$ |
|---|---|---|---|---|
| * | ⊠ ⊠ (0)(7)(8) — — — — — — — — — — — ▮ — — — — — (14)(20) ⊠ <br> 0 1 2 3 4 5 6 7 8 9 10 11 12 13 14 15 16 17 18 19 20 | — — 21 — | — — 21−5 — | — — 3∘2 — |

TABLE XIII
$L = 5$ POSSIBLE WORD SETS

| Number ↓ of → | Words Rejected, $0 \leq N_\Delta \leq 21$ NOTE – BPMs meant below are Δ/B-BPMs. | 0 *or* 21−0 | 1 *or* 21−1 | 2 *or* 21−2 | 3 *or* 21−3 | 4 *or* 21−4 | 5 *or* 21−5 | 6 *or* 21−6 | 7 *or* 21−7 | 8 *or* 21−8 | 9 *or* 21−9 | 10 *or* 21−10 | Line Total |
|---|---|---|---|---|---|---|---|---|---|---|---|---|---|
| Different Word Combinations [showing↓] | | 1 | 21 | 210 | 1,330 | 5,985 | 20,349 | 54,264 | 116,280 | 203,490 | 293,930 | 352,716 | 2,097,152 |
| • shared BPMs among all the $N_\Delta$'s | | — | — | 84 | 836 | 4,596 | 17,228 | 48,422 | 106,932 | 190,384 | 277,656 | 334,606 | 1,961,488 |
| • shared BPMs among the given $N_\Delta$ only | | — | — | 84 | 836 | 4,596 | 17,228 | 48,422 | 106,932 | 190,384 | 277,656 | 334,606 | 1,961,488 |
| *Max Duplication Factor, i.e., number of items sharing a BPM* | | — | — | 4 | 8 | 18 | 36 | 68 | 112 | 156 | 212 | 212 | |
| • unique BPMs among the same scope | | 1 | 21 | 126 | 494 | 1,389 | 3,121 | 5,842 | 9,348 | 13,106 | 16,274 | 18,110 | 135,664 |
| *equivalent percentage* | | 100% | 100% | 60% | ~37% | < 24% | < 16% | < 11% | < 9% | < 7% | < 6% | < 6% | < 7% |

TABLE XIV
BFM/BCM RECOVERY FROM M-BPM

| Given | Iterators | BFM = BFM (M-BPM) | BCM = BCM [M-BPM, BFM (M-BPM)] | Extra Comment |
|---|---|---|---|---|
| M-BPM's $\{{}^{(i)}p_r\}$ | $i$ = *,0...$L$−1; $r$ = 0...$H$−1 | $\{{}^{(i)}f_r\}$ : ${}^{(i=*)}f_r = 1$, ${}^{(i\neq *)}f_0 = \Sigma_r\, {}^{(i-1)}f_r$, ${}^{(i\neq *)}f_{r\neq 0} = {}^{(i-1)}f_{r-1}$ if ${}^{(i)}p_r \neq 0$, else 0 | $\{{}^{(i)}c_r\}$ : ${}^{(i)}c_r = {}^{(i)}p_r$ div ${}^{(i)}f_r$ if ${}^{(i)}p_r \neq 0$ and ${}^{(i)}f_r \neq 0$, else 0 | can be done on the fly with $i$ ascending |

TABLE XV
COMPOSITE ENDEC FRAMEWORK

| Minimal, Ω > 1 ( "one of many to one" ) | General, ω < Ω ( "many of various to one" ) | Maximal, ω = Ω ( "many of one to one" ) |
|---|---|---|
| $\Delta$/B-BPM$_1$ ••• $\Delta$/B-BPM$_\Omega$ | $\Delta$/B-BPM$_1$ ••• $\Delta$/B-BPM$_\Omega$ | $\Delta$/B-BPM$_1$ ••• $\Delta$/B-BPM$_\Omega$ |
| single M-BPM | M-BPM$_1$ ••• M-BPM$_\omega$ | M-BPM$_1$ ••• M-BPM$_\Omega$ |

Anyway, balancing a code via rejection has a sense only in the scope of its initial plot, see Tables IX and XI again.[11]

## IMPLEMENTATION OF A CODE

Given such a decomposed then modified code, whose maps are of known $N_{\rm M} - N_\Delta = N_{\rm B}$, we construct a coding means intended to run an appropriate ENDEC process, either direct, see Table XI yet again, or inverse, see Table XII, depending on what is beneficial in a particular case, see Table XIII, and, typically, choose into the favor of the former option, i.e., with $N_{\rm B}$ out of $N_{\rm M}$ chains kept, when receive $N_\Delta < N_{\rm B}$, and into the favor of the latter, opposite one, i.e., with $N_\Delta$ out of $N_{\rm M}$ chains kept, when receive $N_\Delta > N_{\rm B}$, respectively.

Because a maternal map of such a code "imprints" both its origins losslessly, see Table XIV, there is no need to "imprint" such a bit into the memory of such a coding means.

Moreover, we account for where we deploy such a code, in a stand-alone, or in a composite design, see Table XV.

[11]Facing a code decomposed into a superposition of BPMs, we represent, including visually, see Tables VIII, XI, and XII, its ENDEC process as some superposition, too, collected over as distinct as orthogonal chains of places of indices, $b$'s, in product bins, resulted from the process run with $0 \leq b < N$, where it is $N = N_{\rm M}$ (M-BPM), $N = N_{\rm B}$ (B-BPM), or $N = N_\Delta$ (Δ-BPM), respectively, that pictures both the framework as well as the plot of the code, simultaneously and jointly. (Also for M-BPM, $b \equiv B$, see [3] and [4].)

## SERIALIZATION OF A CODE

Given an ENDEC process, we set out all the content of its (two-dimensional, different-size) product bins regarded to the same letter period, receiving a (unidimensional, uniform-size) runic-like record of such a coherent content, see Table XI yet more again, we use as an extra check for its consistency.

## CONCLUSION

Now, we can handle on quasi base-21 words described by a basic definition, textual, tabular, or mixed, like in [1] and [2], a framework, like in [3], (a bundle of elements of) a plot, like in [4], a match of bin product maps, maternal toward delta or balanced, like in this paper, or a combination thereof.

Balancing a code denoting for (a set of) such words,[12] we decompose, then modify, and then implement it with a coding means running an appropriate ENDEC process.[13]

The latter assumes for a fission–fusion procedure characterized by a couple of numbers, up to which the content of a bin is sourcing into and sourced from, respectively.[14]

## REFERENCES

[1] A. Ivanov, "Base-21 scrambling," ········ @*arXiv*, doi:10.48550/arXiv.yymm.nnnn (bundle), pp. 01–04.

[2] A. Ivanov, "Base-21 word alignment and boundary detection," ········ @*arXiv*, doi:10.48550/arXiv.yymm.nnnn (bundle), pp. 05–08.

[3] A. Ivanov, "Quasi base-21 words," ········ @*arXiv*, doi:10.48550/arXiv.yymm.nnnn (bundle), pp. 09–12.

[4] A. Ivanov, "Quasi base-21 words generated compactly," ········ @*arXiv*, doi:10.48550/arXiv.yymm.nnnn (bundle), pp. 13–16.

[12]Generally speaking, every word of such a code can be either rejected, or repeated, individually, just once and at least once, respectively.

[13]Such a process, see Tables VIII, IX, XI, and XII, is the most critical and the most resource-demanding part, see Table X, of such a means.

[14]We could call those numbers as the fission and fusion factors, agreeably, and thus label the code we consider in this paper, as a $Fi_2Fu_4$ plot code.